\documentclass[conference]{IEEEtran}

\usepackage[utf8]{inputenc}
\usepackage[T1]{fontenc}

\usepackage[none]{hyphenat}

\usepackage[linesnumbered]{algorithm2e}
\usepackage{amsmath, dsfont}
\usepackage{amssymb}
\usepackage{amsthm}
\usepackage[caption=no]{subfig}
\usepackage{tikz}
\usetikzlibrary{calc,decorations,decorations.pathreplacing,positioning,shapes}
\usepackage{pgfplots}
\usepackage{mathtools}
\usepackage{tabularx}
\usepackage{booktabs}
\usepackage{enumitem}

\newtheorem{example}{Example}

\newcommand{\diag}{\ensuremath{\operatorname{diag}}}
\newcommand{\img}{\mathrm{i}}

\SetNlSty{footnotesize}{}{}

\begin{document}
\title{The Role of Multiplicative Complexity in Compiling Low $T$-count Oracle Circuits }

\author{%
  \IEEEauthorblockN{Giulia Meuli$^1$ \quad Mathias Soeken$^1$ \quad Earl Campbell$^2$ \quad Martin Roetteler$^3$ \quad Giovanni De Micheli$^1$}
  \IEEEauthorblockA{%
    $^1$Integrated Systems Laboratory, EPFL, Lausanne, CH \\
    $^2$Department of Physics and Astronomy, University of Sheffield, Sheffield, UK \\
    $^3$Microsoft, Redmond, US
  }
}

\maketitle

\begin{abstract}
  We present a constructive method to create quantum circuits that
  implement oracles
  $|x\rangle|y\rangle|0\rangle^k \mapsto |x\rangle|y \oplus
  f(x)\rangle|0\rangle^k$ for $n$-variable Boolean functions $f$ with
  low $T$-count.  In our method $f$ is given as a 2-regular Boolean
  logic network over the gate basis $\{\land, \oplus, 1\}$.  Our
  construction leads to circuits with a $T$-count that is at most four times the number of AND nodes in the network.
  In addition, we propose a SAT-based method that allows us to trade qubits for $T$ gates, and explore the space/complexity 		trade-off of quantum circuits.

  Our constructive method suggests a new upper bound for the number of $T$ gates
  and ancilla qubits based on the multiplicative complexity $c_\land(f)$ of the
  oracle function $f$, which is the minimum number of AND gates that
  is required to realize $f$ over the gate basis
  $\{\land, \oplus, 1\}$.  There exists a quantum circuit computing
  $f$ with at most $4 c_\land(f)$ $T$ gates using $k = c_\land(f)$
  ancillae.
  Results known for the multiplicative complexity of Boolean functions
  can be transferred. 
  
  We verify our method by comparing it to different state-of-the-art compilers.
  Finally, we present our synthesis results for Boolean functions used in quantum cryptoanalysis. 

\end{abstract}

\section{Introduction}
Quantum computing exploits quantum phenomena such as superposition, entanglement, and interference, in order to provide superior computational capabilities. Many quantum algorithms have been proposed that promise computational speed-ups, e.g., Grover's algorithm~\cite{Grover96} for satisfiability checking, Shor's algorithm~\cite{Shor97} for factoring, and the HHL algorithm~\cite{Harrow09} for solving linear equations.

The computations performed by a quantum computer can be characterized
in terms of unitary matrix operations.  In order to implement them on
a physical quantum device, they must be expressed in terms of the
supported quantum gate library, that is a set
of small unitary matrices.  In fault tolerant quantum computing, the gate library includes
the single-qubit gates $H$ (Hadamard), $S$ (phase), and
$T$ gate, as well as the 2-qubit $\text{\small CNOT}$ gate.  The gates $H$, $S$, and
$\text{\small CNOT}$ are called Clifford gates, and their execution is
significantly less expensive compared to the $T$ gate. Thus, it is
customary to only count the number of $T$ gates ($T$-count) when costing a quantum
computation~\cite{AMMR13}.  

A quantum circuit is a sequence of gates to be executed
on a quantum computer in order to perform a quantum algorithm.
Oracle circuits, that implement the abstract
unitary operation $U_f$ for some Boolean function $f : \{0, 1\}^n \to \{0, 1\}$, play an important role in many quantum algorithms, e.g., Grover's
algorithm~\cite{Grover96}.  In fault tolerant quantum computing, the
quality of an oracle circuit is measured in the $T$-count of
the quantum circuit and the number of helper
qubits (ancillae) required in order to perform
the computation.  

Quantum compilation is the task of decomposing abstract unitary
operations into quantum circuits.
In this paper, we present a constructive compilation algorithm that finds
a quantum circuit for the oracle $U_f$, minimizing the number of $T$ gates. 
The input function $f$ is represented as a 2-regular
Boolean logic network over the gate basis $\{\land, \oplus, 1\}$,
i.e., it consists only of 2-input AND gates, 2-input XOR gates, and
can have constant-1 inputs (0-input gate).  
The construction leads to a quantum circuit consisting of Clifford gates, 
and at most $2\tilde c$ Toffoli
gates, where $\tilde c$ is the number of AND gates in the logic network for
$f$.  

The multiplicative complexity $c_\land(f)$ of a Boolean function $f$
is the minimum number of AND gates that is required to implement it
over the gate basis $\{\land, \oplus, 1\}$~\cite{BPP00}.  The multiplicative complexity of a circuit is the number of AND gates, and therefore an upper bound for the multiplicative complexity of the function it represents.
Our proposed compilation method immediately leads to an upper bound on the number of $T$ gates in a quantum circuit that computes $U_f$: each AND gate is translated into a pair of 2~Toffoli gates that can be realized
using special computation and uncomputation circuits, requiring 4
and 0 $T$ gates respectively~\cite{ Jones13, Gidney18}.  More details about this implementation of the quantum AND gate are given in Section~\ref{quant. circ.}.
It follows that, to implement $U_f$, at most $4c_\land(f)$ $T$ gates and $ c_\land(f)$ extra qubits are required.   
Computing the multiplicative complexity for an
arbitrary Boolean function is intractable~\cite{Find14}. 
Nevertheless, many heuristic algorithms have been proposed to minimize the multiplicative complexity of Boolean circuits~\cite{BP08,testa19,BMP13,CTP19}---almost exclusively motivated by applications in cryptography.
In fact, for some classes of Boolean functions the exact multiplicative complexity
is known, e.g., all Boolean functions with up to 6 inputs~\cite{CTP19} and all symmetric Boolean functions~\cite{BP08}.

We compare the proposed compilation method with a state-of-the-art hierarchical  method based on the Bennett clean-up strategy~\cite{Bennett89}, which aims at reducing the number of $T$ gates and relies on many extra qubits~\cite{SRWM17b}.
To make the comparison fair, we modify the state-of-the-art technique to also implement the quantum AND gate with 4 $T$ gates.
Our experimental comparison demonstrates how our proposed method returns the same number of $T$ gates than the state-of-the-art approach, but  $70\%$ fewer qubits on average. 

In a second evaluation, we compare against the \textit{best-fit} LUT-based compilation method which aims at reducing the number of qubits~\cite{Meuli18}. Our method generates circuits with more qubits, but with $20.4 \times$ fewer $T$ gates on average, for the considered bechmarks. 
The contributions of this work are:
\begin{itemize}[leftmargin=*]
\item we identify the connection between the multiplicative complexity of Boolean functions $c_{\wedge}(f)$ and the number of qubits and $T$-count of a quantum oracle for $f$;
\item we introduce xor-and inverter graphs (XAG) as a suitable multi-level logic representation for the synthesis of quantum oracles;
\item we propose a constructive compilation algorithm that reaches the upper bound of $4 \cdot c_{\wedge}(f)$ $T$ gates;
\item we present a SAT-based reversible pebbling method that acts on XAG networks and allows us to trade qubits for gates, as the number of ancilla qubits may be imposed by the quantum computing device;
\item finally, we provide synthesis results for Boolean functions that could be applied in quantum cryptography.
\end{itemize}

\section{Preliminaries}
\subsection{Quantum states}
While classical computing processes bits, which can be in one of the two ``classical'' logic states 0 and 1, quantum computing acts on qubits, which can be in any superposition of the classical states.

The state $|q\rangle$ of a qubit $q$ is represented by a linear combination of the classical states, $|q\rangle = \alpha |0\rangle + \beta |1\rangle=\binom{\alpha}{\beta}$, with $\alpha, \beta \in \mathds{C}$ and $|\alpha |^2 + |\beta |^2 = 1$. For example, the classical states are represented by $|0\rangle = \binom10$ and $|1\rangle = \binom01$.
A quantum state can be represented by a point on the surface of the Bloch sphere, in which the poles represent the two classical logic states. 
The points on the surface of the sphere are all possible superposed states. For example, all states on the equator of the sphere represent the states with $|\alpha|^2 = |\beta|^2 = \frac 12$, characterized by different angles with respect to the $z$-axis.
While a single-qubit system is characterized by two complex coefficients, a 2-qubit system requires 4 complex coefficients to be represented. In general, to characterize the state of $n$ qubits and to simulate the quantum system behavior on a classical computer, $2^{n}$ complex coefficients are required.

\subsection{Quantum operations}

The state of a qubit can be modified by applying quantum operations. 
All quantum operations that act on $n$ qubits can be represented by $2^n \times 2^n$ unitary matrices.
Quantum devices are operated by means of sets of particular unitary matrices, called quantum gate libraries.
In this work, we target the following operations that are customary when addressing fault tolerant quantum computing: Clifford operations ($H$, $\text{\small CNOT}$, $S$) and the non-Clifford $T$ operation. The matrices of the Clifford+$T$ group are:
\begin{multline}
H = \frac{1}{\sqrt{2}}\begin{pmatrix}
1 & \hfill 1\\
1 & -1\\
\end{pmatrix}, \;
\text{\small CNOT} = \begin{psmallmatrix}
1 & 0 & 0 & 0\\
0 & 1 & 0 & 0\\
0 & 0 & 0 & 1\\
0 & 0 & 1 & 0\\
\end{psmallmatrix}, \\
S = \diag(1, i), \; T= \diag(1, e^{\img\pi / 4}).
\end{multline}
The group also includes the quantum {\small NOT} gate $X = HS^2H = \begin{psmallmatrix} 0&1\\1&0\end{psmallmatrix}$. These gates abstract operations on the physical level. For example, the non-Clifford $T$ gate is injected to the circuit trough a process called magic-state distillation~\cite{BK05}. Many rounds of distillation are required to reach a reasonable error rate (about $10^{-12}$). It has been shown how, when performing  error correction, the $T$ gate results to be more expensive than Clifford gates, independently from the desired error rate~\cite{OGC17}. $T$ gates are generally considered about $50 \times$ more expensive than Clifford gates, in fault tolerant quantum computing.

Computation on a multi-qubit system is modelled by a quantum circuit, which represents a program (set of operations) performed in succession on different qubits. 
Quantum circuits are characterized by three main parameters: (i) number of qubits, (ii) number of gates and (iii) circuit depth. As we already explained, $T$ gates are the most expesive gate in fault tolerant quantum computing. For this reason, we also define: the $T$-count (number of $T$ gates) and the $T$-depth of a circuit.
In this work, we propose an automatic method that minimized the $T$-count, without a significant qubit overhead.

\subsection{Quantum oracle}
\begin{figure}
\centering
\begin{tikzpicture}[scale=1.000000,x=1pt,y=1pt]
\filldraw[color=white] (0.000000, -7.500000) rectangle (92.000000, 67.500000);
\draw[color=black] (0.000000,60.000000) -- (92.000000,60.000000);
\draw[color=black] (0.000000,60.000000) node[left] {$|x\rangle$};
\draw[color=black] (0.000000,45.000000) -- (92.000000,45.000000);
\draw[color=black] (0.000000,45.000000) node[left] {$|0\rangle$};
\draw[color=black] (0.000000,30.000000) node[anchor=mid east] {$\vdots$};
\draw[color=black] (0.000000,15.000000) -- (92.000000,15.000000);
\draw[color=black] (0.000000,15.000000) node[left] {$|0\rangle$};
\draw[color=black] (0.000000,0.000000) -- (92.000000,0.000000);
\draw[color=black] (0.000000,0.000000) node[left] {$|y\rangle$};
\draw (6.000000, 54.000000) -- (14.000000, 66.000000);
\draw (12.000000, 63.000000) node[right] {$\scriptstyle{n}$};
\draw (6.000000, -6.000000) -- (14.000000, 6.000000);
\draw (12.000000, 3.000000) node[right] {$\scriptstyle{m}$};
\draw (46.000000,60.000000) -- (46.000000,0.000000);
\begin{scope}
\draw[fill=white] (46.000000, 30.000000) +(-45.000000:28.284271pt and 50.911688pt) -- +(45.000000:28.284271pt and 50.911688pt) -- +(135.000000:28.284271pt and 50.911688pt) -- +(225.000000:28.284271pt and 50.911688pt) -- cycle;
\clip (46.000000, 30.000000) +(-45.000000:28.284271pt and 50.911688pt) -- +(45.000000:28.284271pt and 50.911688pt) -- +(135.000000:28.284271pt and 50.911688pt) -- +(225.000000:28.284271pt and 50.911688pt) -- cycle;
\draw (46.000000, 30.000000) node {$U_f$};
\end{scope}
\draw (78.000000, 54.000000) -- (86.000000, 66.000000);
\draw (84.000000, 63.000000) node[right] {$\scriptstyle{n}$};
\draw (78.000000, -6.000000) -- (86.000000, 6.000000);
\draw (84.000000, 3.000000) node[right] {$\scriptstyle{m}$};
\draw[color=black] (92.000000,60.000000) node[right] {$|x\rangle$};
\draw[color=black] (92.000000,45.000000) node[right] {$|0\rangle$};
\draw[color=black] (92.000000,30.000000) node[anchor=mid west] {$\vdots$};
\draw[color=black] (92.000000,15.000000) node[right] {$|0\rangle$};
\draw[color=black] (92.000000,0.000000) node[right] {$|y\oplus f(x)\rangle$};
\end{tikzpicture}
\caption{Quantum circuit performing the oracle $U_f$ of a generic multi-input multi-output Boolean function $f$.}
\label{oracle}
\end{figure}
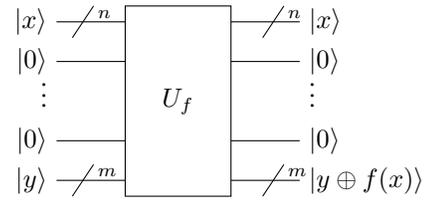

A quantum oracle is defined as a ``black box'' $2^{n+m+k} \times 2^{n+m+k}$ unitary operation $U_f$ performing a multi-output Boolean function $f : \{0, 1\}^n \to \{0, 1\}^m$:
\begin{equation}
  \label{eq:1}
  U_f : |x\rangle|y\rangle|0\rangle^k \mapsto |x\rangle|y \oplus f(x)\rangle|0\rangle^k
\end{equation}
where $|x\rangle$ represents the $n$-qubit input state, $|y\rangle$ the $m$ outputs, and $|0\rangle^k$ are $k$ qubits initialized to the state $|0\rangle$. A generic quantum circuit performing the unitary $U_f$ is shown in Fig.~\ref{oracle}. 
The extra $k$ qubits are called ancillae and are used to store intermediate results for the computation of $f$. They must be restored to $|0\rangle$ as only input and output states must be accessible at the end of the computation. Different automatic clean-up strategies are available in the literature~\cite{Bennett89, Meuli19}, exploring the trade-off between ancillae and operations.
\definecolor{light-gray}{RGB}{233,233,233}

\subsection{Compute/uncompute property of Toffoli gates}\label{quant. circ.}
We consider quantum circuits over the Clifford+$T$ gate set.  Our
construction yields quantum circuits in an intermediate representation
consisting of $X = \text{\small NOT}$, $\text{\small CNOT}$, as well
as Toffoli gates, which either consume or restore an ancilla and can
therefore be implemented as follows~\cite{Gidney18}:
\begin{equation}
  \label{eq:tof-comp}
  \bgroup
  \vcenter{\hbox{\footnotesize
\tikzpicture[scale=1.000000,x=1pt,y=1pt]
\filldraw[color=white] (0.000000, -7.000000) rectangle (155.000000, 35.000000);
\draw[color=black] (0.000000,28.000000) -- (19.500000,28.000000);
\draw[color=black] (57.500000,28.000000) -- (155.000000,28.000000);
\draw[color=black] (0.000000,28.000000) node[left] {$|x_1\rangle$};
\draw[color=black] (0.000000,14.000000) -- (19.500000,14.000000);
\draw[color=black] (57.500000,14.000000) -- (155.000000,14.000000);
\draw[color=black] (0.000000,14.000000) node[left] {$|x_2\rangle$};
\draw[color=black] (0.000000,0.000000) -- (19.500000,0.000000);
\draw[color=black] (57.500000,0.000000) -- (155.000000,0.000000);
\draw[color=black] (0.000000,0.000000) node[left] {$|0\rangle$};
\draw (5.000000,28.000000) -- (5.000000,0.000000);
\filldraw (5.000000, 28.000000) circle(1.500000pt);
\filldraw (5.000000, 14.000000) circle(1.500000pt);
\scope
\draw[fill=white] (5.000000, 0.000000) circle(3.000000pt);
\clip (5.000000, 0.000000) circle(3.000000pt);
\draw (2.000000, 0.000000) -- (8.000000, 0.000000);
\draw (5.000000, -3.000000) -- (5.000000, 3.000000);
\endscope
\draw[color=black] (12.000000,28.000000) node[fill=white,right,minimum height=14.000000pt,minimum width=15.000000pt,inner sep=0pt] {\phantom{$|x_1\rangle$}};
\draw[color=black] (12.000000,28.000000) node[right] {$|x_1\rangle$};
\draw[color=black] (12.000000,14.000000) node[fill=white,right,minimum height=14.000000pt,minimum width=15.000000pt,inner sep=0pt] {\phantom{$|x_2\rangle$}};
\draw[color=black] (12.000000,14.000000) node[right] {$|x_2\rangle$};
\draw[color=black] (12.000000,0.000000) node[fill=white,right,minimum height=14.000000pt,minimum width=15.000000pt,inner sep=0pt] {\phantom{$|x_1x_2\rangle$}};
\draw[color=black] (12.000000,0.000000) node[right] {$|x_1x_2\rangle$};
\draw (38.500000, 14.000000) node {$=$};
\draw[color=black] (65.000000,28.000000) node[fill=white,left,minimum height=14.000000pt,minimum width=15.000000pt,inner sep=0pt] {\phantom{$|x_1\rangle$}};
\draw[color=black] (65.000000,28.000000) node[left] {$|x_1\rangle$};
\draw[color=black] (65.000000,14.000000) node[fill=white,left,minimum height=14.000000pt,minimum width=15.000000pt,inner sep=0pt] {\phantom{$|x_2\rangle$}};
\draw[color=black] (65.000000,14.000000) node[left] {$|x_2\rangle$};
\draw[color=black] (65.000000,0.000000) node[fill=white,left,minimum height=14.000000pt,minimum width=15.000000pt,inner sep=0pt] {\phantom{$|T\rangle$}};
\draw[color=black] (65.000000,0.000000) node[left] {$|T\rangle$};
\draw (72.000000,28.000000) -- (72.000000,0.000000);
\filldraw (72.000000, 28.000000) circle(1.500000pt);
\scope
\draw[fill=white] (72.000000, 0.000000) circle(3.000000pt);
\clip (72.000000, 0.000000) circle(3.000000pt);
\draw (69.000000, 0.000000) -- (75.000000, 0.000000);
\draw (72.000000, -3.000000) -- (72.000000, 3.000000);
\endscope
\draw (82.000000,14.000000) -- (82.000000,0.000000);
\filldraw (82.000000, 14.000000) circle(1.500000pt);
\scope
\draw[fill=white] (82.000000, 0.000000) circle(3.000000pt);
\clip (82.000000, 0.000000) circle(3.000000pt);
\draw (79.000000, 0.000000) -- (85.000000, 0.000000);
\draw (82.000000, -3.000000) -- (82.000000, 3.000000);
\endscope
\draw (92.000000,28.000000) -- (92.000000,0.000000);
\filldraw (92.000000, 0.000000) circle(1.500000pt);
\scope
\draw[fill=white] (92.000000, 28.000000) circle(3.000000pt);
\clip (92.000000, 28.000000) circle(3.000000pt);
\draw (89.000000, 28.000000) -- (95.000000, 28.000000);
\draw (92.000000, 25.000000) -- (92.000000, 31.000000);
\endscope
\scope
\draw[fill=white] (92.000000, 14.000000) circle(3.000000pt);
\clip (92.000000, 14.000000) circle(3.000000pt);
\draw (89.000000, 14.000000) -- (95.000000, 14.000000);
\draw (92.000000, 11.000000) -- (92.000000, 17.000000);
\endscope
\scope
\draw[fill=white] (105.000000, 28.000000) +(-45.000000:8.485281pt and 8.485281pt) -- +(45.000000:8.485281pt and 8.485281pt) -- +(135.000000:8.485281pt and 8.485281pt) -- +(225.000000:8.485281pt and 8.485281pt) -- cycle;
\clip (105.000000, 28.000000) +(-45.000000:8.485281pt and 8.485281pt) -- +(45.000000:8.485281pt and 8.485281pt) -- +(135.000000:8.485281pt and 8.485281pt) -- +(225.000000:8.485281pt and 8.485281pt) -- cycle;
\draw (105.000000, 28.000000) node {{$T^\dagger$}};
\endscope
\scope
\draw[fill=white] (105.000000, 14.000000) +(-45.000000:8.485281pt and 8.485281pt) -- +(45.000000:8.485281pt and 8.485281pt) -- +(135.000000:8.485281pt and 8.485281pt) -- +(225.000000:8.485281pt and 8.485281pt) -- cycle;
\clip (105.000000, 14.000000) +(-45.000000:8.485281pt and 8.485281pt) -- +(45.000000:8.485281pt and 8.485281pt) -- +(135.000000:8.485281pt and 8.485281pt) -- +(225.000000:8.485281pt and 8.485281pt) -- cycle;
\draw (105.000000, 14.000000) node {{$T^\dagger$}};
\endscope
\scope
\draw[fill=white] (105.000000, -0.000000) +(-45.000000:8.485281pt and 8.485281pt) -- +(45.000000:8.485281pt and 8.485281pt) -- +(135.000000:8.485281pt and 8.485281pt) -- +(225.000000:8.485281pt and 8.485281pt) -- cycle;
\clip (105.000000, -0.000000) +(-45.000000:8.485281pt and 8.485281pt) -- +(45.000000:8.485281pt and 8.485281pt) -- +(135.000000:8.485281pt and 8.485281pt) -- +(225.000000:8.485281pt and 8.485281pt) -- cycle;
\draw (105.000000, -0.000000) node {{$T$}};
\endscope
\draw (118.000000,28.000000) -- (118.000000,0.000000);
\filldraw (118.000000, 0.000000) circle(1.500000pt);
\scope
\draw[fill=white] (118.000000, 28.000000) circle(3.000000pt);
\clip (118.000000, 28.000000) circle(3.000000pt);
\draw (115.000000, 28.000000) -- (121.000000, 28.000000);
\draw (118.000000, 25.000000) -- (118.000000, 31.000000);
\endscope
\scope
\draw[fill=white] (118.000000, 14.000000) circle(3.000000pt);
\clip (118.000000, 14.000000) circle(3.000000pt);
\draw (115.000000, 14.000000) -- (121.000000, 14.000000);
\draw (118.000000, 11.000000) -- (118.000000, 17.000000);
\endscope
\scope
\draw[fill=white] (131.000000, -0.000000) +(-45.000000:8.485281pt and 8.485281pt) -- +(45.000000:8.485281pt and 8.485281pt) -- +(135.000000:8.485281pt and 8.485281pt) -- +(225.000000:8.485281pt and 8.485281pt) -- cycle;
\clip (131.000000, -0.000000) +(-45.000000:8.485281pt and 8.485281pt) -- +(45.000000:8.485281pt and 8.485281pt) -- +(135.000000:8.485281pt and 8.485281pt) -- +(225.000000:8.485281pt and 8.485281pt) -- cycle;
\draw (131.000000, -0.000000) node {$H$};
\endscope
\scope
\draw[fill=white] (147.000000, -0.000000) +(-45.000000:8.485281pt and 8.485281pt) -- +(45.000000:8.485281pt and 8.485281pt) -- +(135.000000:8.485281pt and 8.485281pt) -- +(225.000000:8.485281pt and 8.485281pt) -- cycle;
\clip (147.000000, -0.000000) +(-45.000000:8.485281pt and 8.485281pt) -- +(45.000000:8.485281pt and 8.485281pt) -- +(135.000000:8.485281pt and 8.485281pt) -- +(225.000000:8.485281pt and 8.485281pt) -- cycle;
\draw (147.000000, -0.000000) node {{$S$}};
\endscope
\draw[color=black] (155.000000,28.000000) node[right] {$|x_1\rangle$};
\draw[color=black] (155.000000,14.000000) node[right] {$|x_2\rangle$};
\draw[color=black] (155.000000,0.000000) node[right] {$|x_1x_2\rangle$};
\endtikzpicture}}
  \egroup
\end{equation}
\begin{equation}
  \label{eq:tof-uncomp}
  \bgroup
  \vcenter{\hbox{\begin{tikzpicture}[scale=1.000000,x=1pt,y=1pt]
\filldraw[color=white] (0.000000, -7.000000) rectangle (147.000000, 35.000000);
\draw[color=black] (0.000000,28.000000) -- (31.500000,28.000000);
\draw[color=black] (85.500000,28.000000) -- (147.000000,28.000000);
\draw[color=black] (0.000000,28.000000) node[left] {$|x_1\rangle$};
\draw[color=black] (0.000000,14.000000) -- (31.500000,14.000000);
\draw[color=black] (85.500000,14.000000) -- (147.000000,14.000000);
\draw[color=black] (0.000000,14.000000) node[left] {$|x_2\rangle$};
\draw[color=black] (0.000000,0.000000) -- (31.500000,0.000000);
\draw[color=black] (85.500000,0.000000) -- (135.000000,0.000000);
\draw[color=black] (135.000000,-0.500000) -- (147.000000,-0.500000);
\draw[color=black] (135.000000,0.500000) -- (147.000000,0.500000);
\draw[color=black] (0.000000,0.000000) node[left] {$|x_1x_2\rangle$};
\draw (9.000000,28.000000) -- (9.000000,0.000000);
\filldraw (9.000000, 28.000000) circle(1.500000pt);
\filldraw (9.000000, 14.000000) circle(1.500000pt);
\begin{scope}
\draw[fill=white] (9.000000, 0.000000) circle(3.000000pt);
\clip (9.000000, 0.000000) circle(3.000000pt);
\draw (6.000000, 0.000000) -- (12.000000, 0.000000);
\draw (9.000000, -3.000000) -- (9.000000, 3.000000);
\end{scope}
\draw[color=black] (24.000000,28.000000) node[fill=white,right,minimum height=14.000000pt,minimum width=15.000000pt,inner sep=0pt] {\phantom{$|x_1\rangle$}};
\draw[color=black] (24.000000,28.000000) node[right] {$|x_1\rangle$};
\draw[color=black] (24.000000,14.000000) node[fill=white,right,minimum height=14.000000pt,minimum width=15.000000pt,inner sep=0pt] {\phantom{$|x_2\rangle$}};
\draw[color=black] (24.000000,14.000000) node[right] {$|x_2\rangle$};
\draw[color=black] (24.000000,0.000000) node[fill=white,right,minimum height=14.000000pt,minimum width=15.000000pt,inner sep=0pt] {\phantom{$|0\rangle$}};
\draw[color=black] (24.000000,0.000000) node[right] {$|0\rangle$};
\draw[fill=white,color=white] (51.000000, -6.000000) rectangle (66.000000, 34.000000);
\draw (58.500000, 14.000000) node {$=$};
\draw[color=black] (93.000000,28.000000) node[fill=white,left,minimum height=14.000000pt,minimum width=15.000000pt,inner sep=0pt] {\phantom{$|x_1\rangle$}};
\draw[color=black] (93.000000,28.000000) node[left] {$|x_1\rangle$};
\draw[color=black] (93.000000,14.000000) node[fill=white,left,minimum height=14.000000pt,minimum width=15.000000pt,inner sep=0pt] {\phantom{$|x_2\rangle$}};
\draw[color=black] (93.000000,14.000000) node[left] {$|x_2\rangle$};
\draw[color=black] (93.000000,0.000000) node[fill=white,left,minimum height=14.000000pt,minimum width=15.000000pt,inner sep=0pt] {\phantom{$|x_1x_2\rangle$}};
\draw[color=black] (93.000000,0.000000) node[left] {$|x_1x_2\rangle$};
\begin{scope}
\draw[fill=white] (111.000000, -0.000000) +(-45.000000:8.485281pt and 8.485281pt) -- +(45.000000:8.485281pt and 8.485281pt) -- +(135.000000:8.485281pt and 8.485281pt) -- +(225.000000:8.485281pt and 8.485281pt) -- cycle;
\clip (111.000000, -0.000000) +(-45.000000:8.485281pt and 8.485281pt) -- +(45.000000:8.485281pt and 8.485281pt) -- +(135.000000:8.485281pt and 8.485281pt) -- +(225.000000:8.485281pt and 8.485281pt) -- cycle;
\draw (111.000000, -0.000000) node {$H$};
\end{scope}
\draw (135.000000,28.000000) -- (135.000000,14.000000);
\draw (134.500000,14.000000) -- (134.500000,0.000000);
\draw (135.500000,14.000000) -- (135.500000,0.000000);
\begin{scope}
\draw[fill=white] (135.000000, 14.000000) +(-45.000000:8.485281pt and 8.485281pt) -- +(45.000000:8.485281pt and 8.485281pt) -- +(135.000000:8.485281pt and 8.485281pt) -- +(225.000000:8.485281pt and 8.485281pt) -- cycle;
\clip (135.000000, 14.000000) +(-45.000000:8.485281pt and 8.485281pt) -- +(45.000000:8.485281pt and 8.485281pt) -- +(135.000000:8.485281pt and 8.485281pt) -- +(225.000000:8.485281pt and 8.485281pt) -- cycle;
\draw (135.000000, 14.000000) node {$Z$};
\end{scope}
\filldraw (135.000000, 0.000000) circle(1.500000pt);
\filldraw (135.000000, 28.000000) circle(1.500000pt);
\draw[fill=white] (129.000000, -6.000000) rectangle (141.000000, 6.000000);
\draw[very thin] (135.000000, 0.600000) arc (90:150:6.000000pt);
\draw[very thin] (135.000000, 0.600000) arc (90:30:6.000000pt);
\draw[->,>=stealth] (135.000000, -5.400000) -- +(80:10.392305pt);
\draw[color=black] (147.000000,28.000000) node[right] {$|x_1\rangle$};
\draw[color=black] (147.000000,14.000000) node[right] {$|x_2\rangle$};
\end{tikzpicture}}}
  \egroup
\end{equation}
In~\eqref{eq:tof-comp} the state $|T\rangle = TH|0\rangle$ can be
applied using magic-state distillation (see, e.g., \cite{BK05}).  The
other three $T$ gates in the circuit can be realized using magic-state
distillation and gate teleportation (see, e.g., \cite{GC99}).

In~\eqref{eq:tof-uncomp} the original state is recovered by measuring the third qubit and applying a controlled-$Z$ rotation (can be realized using 2 $H$ gates and a {\small CNOT} gate) to correct a $-1$ phase due to the measurement back-action. In fact, we have
\begin{equation}
  \label{eq:eq}
  \bgroup
  \vcenter{\hbox{\begin{tikzpicture}[scale=1.000000,x=1pt,y=1pt]
\filldraw[color=white] (0.000000, -7.000000) rectangle (117.000000, 35.000000);
\draw[color=black] (0.000000,28.000000) -- (117.000000,28.000000);
\draw[color=black] (0.000000,28.000000) node[left] {$|x_1\rangle$};
\draw[color=black] (0.000000,14.000000) -- (117.000000,14.000000);
\draw[color=black] (0.000000,14.000000) node[left] {$|x_2\rangle$};
\draw[color=black] (0.000000,0.000000) -- (117.000000,0.000000);
\draw[color=black] (0.000000,0.000000) node[left] {$|x_1x_2\rangle$};
\draw (9.000000,28.000000) -- (9.000000,0.000000);
\filldraw (9.000000, 28.000000) circle(1.500000pt);
\filldraw (9.000000, 14.000000) circle(1.500000pt);
\begin{scope}
\draw[fill=white] (9.000000, 0.000000) circle(3.000000pt);
\clip (9.000000, 0.000000) circle(3.000000pt);
\draw (6.000000, 0.000000) -- (12.000000, 0.000000);
\draw (9.000000, -3.000000) -- (9.000000, 3.000000);
\end{scope}
\draw[fill=white,color=white] (24.000000, -6.000000) rectangle (39.000000, 34.000000);
\draw (31.500000, 14.000000) node {$=$};
\begin{scope}
\draw[fill=white] (57.000000, -0.000000) +(-45.000000:8.485281pt and 8.485281pt) -- +(45.000000:8.485281pt and 8.485281pt) -- +(135.000000:8.485281pt and 8.485281pt) -- +(225.000000:8.485281pt and 8.485281pt) -- cycle;
\clip (57.000000, -0.000000) +(-45.000000:8.485281pt and 8.485281pt) -- +(45.000000:8.485281pt and 8.485281pt) -- +(135.000000:8.485281pt and 8.485281pt) -- +(225.000000:8.485281pt and 8.485281pt) -- cycle;
\draw (57.000000, -0.000000) node {$H$};
\end{scope}
\draw (81.000000,28.000000) -- (81.000000,0.000000);
\begin{scope}
\draw[fill=white] (81.000000, -0.000000) +(-45.000000:8.485281pt and 8.485281pt) -- +(45.000000:8.485281pt and 8.485281pt) -- +(135.000000:8.485281pt and 8.485281pt) -- +(225.000000:8.485281pt and 8.485281pt) -- cycle;
\clip (81.000000, -0.000000) +(-45.000000:8.485281pt and 8.485281pt) -- +(45.000000:8.485281pt and 8.485281pt) -- +(135.000000:8.485281pt and 8.485281pt) -- +(225.000000:8.485281pt and 8.485281pt) -- cycle;
\draw (81.000000, -0.000000) node {$Z$};
\end{scope}
\filldraw (81.000000, 28.000000) circle(1.500000pt);
\filldraw (81.000000, 14.000000) circle(1.500000pt);
\begin{scope}
\draw[fill=white] (105.000000, -0.000000) +(-45.000000:8.485281pt and 8.485281pt) -- +(45.000000:8.485281pt and 8.485281pt) -- +(135.000000:8.485281pt and 8.485281pt) -- +(225.000000:8.485281pt and 8.485281pt) -- cycle;
\clip (105.000000, -0.000000) +(-45.000000:8.485281pt and 8.485281pt) -- +(45.000000:8.485281pt and 8.485281pt) -- +(135.000000:8.485281pt and 8.485281pt) -- +(225.000000:8.485281pt and 8.485281pt) -- cycle;
\draw (105.000000, -0.000000) node {$H$};
\end{scope}
\draw[color=black] (117.000000,28.000000) node[right] {$|x_1\rangle$};
\draw[color=black] (117.000000,14.000000) node[right] {$|x_2\rangle$};
\draw[color=black] (117.000000,0.000000) node[right] {$|0\rangle$};
\end{tikzpicture}}}
  \egroup
\end{equation}
and
\begin{equation}
  \label{eq:eq}
  \bgroup
  \vcenter{\hbox{
\begin{tikzpicture}[scale=1.000000,x=1pt,y=1pt]
\filldraw[color=white] (0.000000, -7.000000) rectangle (24.000000, 35.000000);
\draw[color=black] (0.000000,28.000000) -- (24.000000,28.000000);
\draw[color=black] (0.000000,28.000000) node[left] {$|x_1\rangle$};
\draw[color=black] (0.000000,14.000000) -- (24.000000,14.000000);
\draw[color=black] (0.000000,14.000000) node[left] {$|x_2\rangle$};
\draw[color=black] (0.000000,0.000000) -- (24.000000,0.000000);
\draw[color=black] (0.000000,0.000000) node[left] {$|x_3\rangle$};
\draw (12.000000,28.000000) -- (12.000000,0.000000);
\begin{scope}
\draw[fill=white] (12.000000, -0.000000) +(-45.000000:8.485281pt and 8.485281pt) -- +(45.000000:8.485281pt and 8.485281pt) -- +(135.000000:8.485281pt and 8.485281pt) -- +(225.000000:8.485281pt and 8.485281pt) -- cycle;
\clip (12.000000, -0.000000) +(-45.000000:8.485281pt and 8.485281pt) -- +(45.000000:8.485281pt and 8.485281pt) -- +(135.000000:8.485281pt and 8.485281pt) -- +(225.000000:8.485281pt and 8.485281pt) -- cycle;
\draw (12.000000, -0.000000) node {$Z$};
\end{scope}
\filldraw (12.000000, 28.000000) circle(1.500000pt);
\filldraw (12.000000, 14.000000) circle(1.500000pt);
\filldraw[color=white,fill=white] (24.000000,-3.500000) rectangle (28.000000,31.500000);
\draw[decorate,decoration={brace,mirror,amplitude = 4.000000pt},very thick] (24.000000,-3.500000) -- (24.000000,31.500000);
\draw[color=black] (28.000000,14.000000) node[right] {$(-1)^{x_1x_2x_3}|x_1x_2x_3\rangle$};
\end{tikzpicture}}}.
  \egroup
\end{equation}
If measuring the third qubit yields a $1$ as a result, we must compensate the introduced $-1$ phase. We can do so by applying a controlled-$Z$ gate, where $Z = HXH$.
 
Computing the Toffoli gate (logical-AND) requires 4 $T$ gates (see~\eqref{eq:tof-comp}), while uncomputing only requires Clifford gates (see~\eqref{eq:tof-uncomp}). This asymmetry is due to the fact that measurement is not reversible. It has been proven that a Toffoli gate cannot be computed with less than 4 $T$ gates~\cite{HC17}.
Our proposed compilation approach is designed to maximally take advantage of this implementation, to reduce the number of $T$ gates of the final circuit.

\subsection{Logic networks}
In this work, we are considering logic networks over the gate basis
$\{\land, \oplus, 1\}$.  In order to simplify our compiling algorithm,
we allow our logic networks to have inverters.  We can then propagate
all uses of the constant 1 input to the outputs, since
$1 \oplus x = \bar x$, and $1 \land x = x$, without increasing the
number of AND gates.  Consequently, the multiplicative complexity of a
Boolean function over the gate set $\{\land, \oplus, 1\}$ is the
equivalent to the multiplicative complexity over the gate set
$\{\land, \oplus, \neg\}$~\cite{BPP00}.  We use $\bar x$ to denote the
Boolean complement of $x = 1 - x$, and define $x^0 = \bar x$ and
$x^1 = x$.

We model a logic network for an $n$-variable Boolean function with
inputs $x_1, \dots, x_n$ as a Boolean chain with steps
\begin{equation}
  \label{eq:chain}
  x_i = x_{j(i)} \oplus x_{k(i)} \qquad\text{or}\qquad
  x_i = x_{j(i)}^{p(i)} \land x_{k(i)}^{q(i)},
\end{equation}
for $n < i \le n + r$, depending on whether the step computes the XOR
or the AND operation, where $r$ is the number of steps. The constant values $1 \le j(i) < k(i) < i$
point to input or previous steps in the chain, and in the case of an
AND gate Boolean constants $p(i)$ and $q(i)$ are used to possibly complement
the gate's fan-in.  The function value is computed by the last step
$f = x_{n+r}^p$, which may be complemented.  We write
$\circ_i = \land$, if step $i$ computes an AND gate, and
$\circ_i = \oplus$, if step $i$ computes an XOR gate.  We define
$x_0 = 0$, i.e., logic networks with no inputs and no steps represent
the constant functions.  The number of AND gates in the logic network
is $\tilde c = |\{i \mid \circ_i = \land\}|$, which is an upper bound
of the multiplicative complexity of the Boolean function it realizes.

\begin{example}
  \label{ex:maj-network}
  The majority-of-three function
  $\langle x_1x_2x_3\rangle = x_1x_2 \lor x_1x_3 \lor x_2x_3$ can be
  realized by the logic network
  \[
    \begin{aligned}
      x_4 &= x_1 \oplus x_2, \qquad
      x_5 &= x_2 \oplus x_3, \\
      x_6 &= x_4 \land x_5, \qquad
      x_7 &= x_2 \oplus x_6,
    \end{aligned}
  \]
  with $\tilde c = 1$.
\end{example}

\section{Compilation Algorithm}\label{comp. alg.}
In this section, we describe a synthesis algorithm that, given a logic
network computing an $n$-variable Boolean function $f(x)$, finds a
quantum circuit that implements the unitary operation
\begin{equation}
  U_f : |x\rangle|y\rangle|0\rangle^{\tilde c} \mapsto |x\rangle|y \oplus f(x)\rangle|0\rangle^{\tilde c}
\end{equation}
using $4 \tilde c$ $T$ gates.  For the sake of clarity, we describe
the single-output function case. Our actual implementation is a generalization of the described algorithm that also supports multi-output functions.

The key insight is that each AND gate in the logic network is driven
by two multi-input parity functions of variables which are either
inputs or AND steps in the logic network.  Note that the arity of this
multi-input parity functions might be 1.  This is the case when the
immediate input to an AND gate is a primary input or another AND gate
itself.  Formally, let the linear transitive fan-in of a node $x_i$ in
a logic network be defined using the recursive function
\newcommand{\ltfi}{\ensuremath{\operatorname{ltfi}}}
\begin{equation}
  \label{eq:ltfi}
  \begin{small}
    \ltfi(x_i) =
    \begin{cases}
      \{x_i\} & \text{if $i \le n$ or $\circ_i = \land$,} \\
      \ltfi(x_{j(i)}) \mathbin{\triangle} \ltfi(x_{k(i)}) & \text{otherwise,}
    \end{cases}
  \end{small}
\end{equation}
where `$\triangle$' denotes the symmetric difference of two sets.  It
is easy to see that all elements in $\ltfi(x_i)$ are either inputs or
steps that compute an AND gate.

\begin{example}
  For the network in Example~\ref{ex:maj-network}, we have
  \[
    \begin{aligned}
      \ltfi(x_4) &= \{x_1, x_2\} \\
      \ltfi(x_5) &= \{x_2, x_3\} \\
      \ltfi(x_6) &= \{x_6\} \\
      \ltfi(x_7) &= \{x_2, x_6\}.
    \end{aligned}
  \]
\end{example}

\begin{figure}[t!]
  \centering
  \subfloat[AND step in logic network]{%
    \begin{tikzpicture}[font=\footnotesize]
      \node[inner sep=1pt,draw,circle] (xi) {$\land$};
      \draw (xi) -- ++(up:12pt) node[above,inner sep=1pt] {$x_i$};
      \node[draw,cloud,cloud puffs=7,inner sep=2pt,minimum width=.8cm] (cloud1) at ($(xi)+(225:1cm)$) {$\oplus$};
      \node[draw,cloud,cloud puffs=7,inner sep=2pt,minimum width=.8cm] (cloud2) at ($(xi)+(315:1cm)$) {$\oplus$};
      \draw (xi) -- (cloud1) (xi) -- (cloud2);
      \node[below,inner sep=3pt] at (cloud1.south) {$\ltfi(x_{j(i)})$};
      \node[below,inner sep=3pt] at (cloud2.south) {$\ltfi(x_{k(i)})$};
      \fill[gray!50!white] ($(xi)!.4!(cloud1)$) circle (2pt) node[left,yshift=3pt,inner sep=1.5pt,black,font=\scriptsize] {$p(i)$};
      \fill[gray!50!white] ($(xi)!.4!(cloud2)$) circle (2pt) node[right,yshift=3pt,inner sep=1.5pt,black,font=\scriptsize] {$q(i)$};
    \end{tikzpicture}}
  \hfil
  \subfloat[Quantum circuit construction for AND step]{%
    \footnotesize
\tikzpicture[scale=0.600000,x=1pt,y=1pt]
\filldraw[color=white] (0.000000, -6.500000) rectangle (89.000000, 136.500000);
\draw[color=black] (0.000000,130.000000) -- (89.000000,130.000000);
\draw[color=black] (0.000000,117.000000) -- (89.000000,117.000000);
\draw[color=black] (0.000000,104.000000) node[left] {${}$};
\draw[color=black] (0.000000,91.000000) -- (89.000000,91.000000);
\draw[color=black] (0.000000,78.000000) -- (89.000000,78.000000);
\filldraw[color=white,fill=white] (0.000000,74.750000) rectangle (-4.000000,133.250000);
\draw[decorate,decoration={brace,amplitude = 4.000000pt},very thick] (0.000000,74.750000) -- (0.000000,133.250000);
\draw[color=black] (-4.000000,104.000000) node[left] {${\operatorname{ltfi}(x_{j(i)})}$};
\draw[color=black] (0.000000,65.000000) -- (89.000000,65.000000);
\draw[color=black] (0.000000,52.000000) -- (89.000000,52.000000);
\draw[color=black] (0.000000,39.000000) node[left] {${}$};
\draw[color=black] (0.000000,26.000000) -- (89.000000,26.000000);
\draw[color=black] (0.000000,13.000000) -- (89.000000,13.000000);
\filldraw[color=white,fill=white] (0.000000,9.750000) rectangle (-4.000000,68.250000);
\draw[decorate,decoration={brace,amplitude = 4.000000pt},very thick] (0.000000,9.750000) -- (0.000000,68.250000);
\draw[color=black] (-4.000000,39.000000) node[left] {${\operatorname{ltfi}(x_{k(i)})}$};
\draw[color=black] (0.000000,0.000000) -- (89.000000,0.000000);
\draw[color=black] (0.000000,0.000000) node[left] {$|0\rangle$};
\draw (4.000000,130.000000) -- (4.000000,78.000000);
\filldraw (4.000000, 130.000000) circle(1.500000pt);
\scope
\draw[fill=white] (4.000000, 78.000000) circle(3.000000pt);
\clip (4.000000, 78.000000) circle(3.000000pt);
\draw (1.000000, 78.000000) -- (7.000000, 78.000000);
\draw (4.000000, 75.000000) -- (4.000000, 81.000000);
\endscope
\draw (4.000000,65.000000) -- (4.000000,13.000000);
\filldraw (4.000000, 65.000000) circle(1.500000pt);
\scope
\draw[fill=white] (4.000000, 13.000000) circle(3.000000pt);
\clip (4.000000, 13.000000) circle(3.000000pt);
\draw (1.000000, 13.000000) -- (7.000000, 13.000000);
\draw (4.000000, 10.000000) -- (4.000000, 16.000000);
\endscope
\draw (12.000000,117.000000) -- (12.000000,78.000000);
\filldraw (12.000000, 117.000000) circle(1.500000pt);
\scope
\draw[fill=white] (12.000000, 78.000000) circle(3.000000pt);
\clip (12.000000, 78.000000) circle(3.000000pt);
\draw (9.000000, 78.000000) -- (15.000000, 78.000000);
\draw (12.000000, 75.000000) -- (12.000000, 81.000000);
\endscope
\draw (12.000000,52.000000) -- (12.000000,13.000000);
\filldraw (12.000000, 52.000000) circle(1.500000pt);
\scope
\draw[fill=white] (12.000000, 13.000000) circle(3.000000pt);
\clip (12.000000, 13.000000) circle(3.000000pt);
\draw (9.000000, 13.000000) -- (15.000000, 13.000000);
\draw (12.000000, 10.000000) -- (12.000000, 16.000000);
\endscope
\draw (20.000000,91.000000) -- (20.000000,78.000000);
\filldraw (20.000000, 91.000000) circle(1.500000pt);
\scope
\draw[fill=white] (20.000000, 78.000000) circle(3.000000pt);
\clip (20.000000, 78.000000) circle(3.000000pt);
\draw (17.000000, 78.000000) -- (23.000000, 78.000000);
\draw (20.000000, 75.000000) -- (20.000000, 81.000000);
\endscope
\draw (20.000000,26.000000) -- (20.000000,13.000000);
\filldraw (20.000000, 26.000000) circle(1.500000pt);
\scope
\draw[fill=white] (20.000000, 13.000000) circle(3.000000pt);
\clip (20.000000, 13.000000) circle(3.000000pt);
\draw (17.000000, 13.000000) -- (23.000000, 13.000000);
\draw (20.000000, 10.000000) -- (20.000000, 16.000000);
\endscope
\draw[color=black] (32.500000, 78.000000) node [inner sep=.5pt,fill=white] {$t_1$};
\draw[color=black] (32.500000, 13.000000) node [inner sep=.5pt,fill=white] {$t_2$};
\scope
\draw[gray!50!white,fill=white] (45.000000, 78.000000) circle(3.000000pt);
\clip (45.000000, 78.000000) circle(3.000000pt);
\draw[gray!50!white] (42.000000, 78.000000) -- (48.000000, 78.000000);
\draw[gray!50!white] (45.000000, 75.000000) -- (45.000000, 81.000000);
\endscope
\scope
\draw[gray!50!white,fill=white] (45.000000, 13.000000) circle(3.000000pt);
\clip (45.000000, 13.000000) circle(3.000000pt);
\draw[gray!50!white] (42.000000, 13.000000) -- (48.000000, 13.000000);
\draw[gray!50!white] (45.000000, 10.000000) -- (45.000000, 16.000000);
\endscope
\draw (53.000000,78.000000) -- (53.000000,0.000000);
\filldraw (53.000000, 78.000000) circle(1.500000pt);
\filldraw (53.000000, 13.000000) circle(1.500000pt);
\scope
\draw[fill=white] (53.000000, 0.000000) circle(3.000000pt);
\clip (53.000000, 0.000000) circle(3.000000pt);
\draw (50.000000, 0.000000) -- (56.000000, 0.000000);
\draw (53.000000, -3.000000) -- (53.000000, 3.000000);
\endscope
\scope
\draw[gray!50!white,fill=white] (61.000000, 13.000000) circle(3.000000pt);
\clip (61.000000, 13.000000) circle(3.000000pt);
\draw[gray!50!white] (58.000000, 13.000000) -- (64.000000, 13.000000);
\draw[gray!50!white] (61.000000, 10.000000) -- (61.000000, 16.000000);
\endscope
\scope
\draw[gray!50!white,fill=white] (61.000000, 78.000000) circle(3.000000pt);
\clip (61.000000, 78.000000) circle(3.000000pt);
\draw[gray!50!white] (58.000000, 78.000000) -- (64.000000, 78.000000);
\draw[gray!50!white] (61.000000, 75.000000) -- (61.000000, 81.000000);
\endscope
\draw (69.000000,26.000000) -- (69.000000,13.000000);
\filldraw (69.000000, 26.000000) circle(1.500000pt);
\scope
\draw[fill=white] (69.000000, 13.000000) circle(3.000000pt);
\clip (69.000000, 13.000000) circle(3.000000pt);
\draw (66.000000, 13.000000) -- (72.000000, 13.000000);
\draw (69.000000, 10.000000) -- (69.000000, 16.000000);
\endscope
\draw (69.000000,91.000000) -- (69.000000,78.000000);
\filldraw (69.000000, 91.000000) circle(1.500000pt);
\scope
\draw[fill=white] (69.000000, 78.000000) circle(3.000000pt);
\clip (69.000000, 78.000000) circle(3.000000pt);
\draw (66.000000, 78.000000) -- (72.000000, 78.000000);
\draw (69.000000, 75.000000) -- (69.000000, 81.000000);
\endscope
\draw (77.000000,52.000000) -- (77.000000,13.000000);
\filldraw (77.000000, 52.000000) circle(1.500000pt);
\scope
\draw[fill=white] (77.000000, 13.000000) circle(3.000000pt);
\clip (77.000000, 13.000000) circle(3.000000pt);
\draw (74.000000, 13.000000) -- (80.000000, 13.000000);
\draw (77.000000, 10.000000) -- (77.000000, 16.000000);
\endscope
\draw (77.000000,117.000000) -- (77.000000,78.000000);
\filldraw (77.000000, 117.000000) circle(1.500000pt);
\scope
\draw[fill=white] (77.000000, 78.000000) circle(3.000000pt);
\clip (77.000000, 78.000000) circle(3.000000pt);
\draw (74.000000, 78.000000) -- (80.000000, 78.000000);
\draw (77.000000, 75.000000) -- (77.000000, 81.000000);
\endscope
\draw (85.000000,65.000000) -- (85.000000,13.000000);
\filldraw (85.000000, 65.000000) circle(1.500000pt);
\scope
\draw[fill=white] (85.000000, 13.000000) circle(3.000000pt);
\clip (85.000000, 13.000000) circle(3.000000pt);
\draw (82.000000, 13.000000) -- (88.000000, 13.000000);
\draw (85.000000, 10.000000) -- (85.000000, 16.000000);
\endscope
\draw (85.000000,130.000000) -- (85.000000,78.000000);
\filldraw (85.000000, 130.000000) circle(1.500000pt);
\scope
\draw[fill=white] (85.000000, 78.000000) circle(3.000000pt);
\clip (85.000000, 78.000000) circle(3.000000pt);
\draw (82.000000, 78.000000) -- (88.000000, 78.000000);
\draw (85.000000, 75.000000) -- (85.000000, 81.000000);
\endscope
\draw[color=black] (89.000000,104.000000) node[anchor=mid west] {$\vdots$};
\draw[color=black] (89.000000,39.000000) node[anchor=mid west] {$\vdots$};
\draw[color=black] (89.000000,0.000000) node[right] {$|x_i\rangle$};
\endtikzpicture
  }
  \caption{Illustration of the general idea in which the fan-in nodes of an
    AND gate are considered as large XOR gates.  These can be computed
    and uncomputed in-place in the quantum circuit using CNOT gates.}
  \label{fig:and-step}
\end{figure}
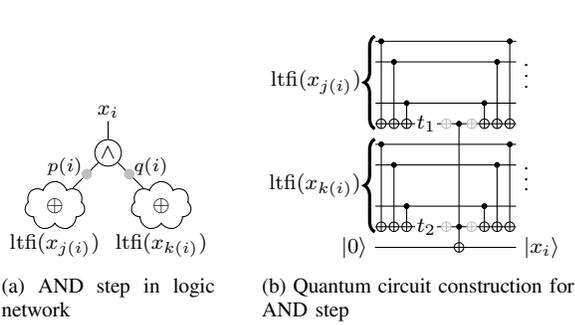

Algorithm~\ref{alg:heuristic} is based on this
idea.  Lines~\ref{alg:main-begin}--\ref{alg:main-end} show that the
algorithm first computes all intermediate signals using the function
`\emph{compute}' then copies the output to the qubit $y$, before
restoring all ancilla qubits to $|0\rangle$ by uncomputing
`\emph{compute}'.  The function computes in
lines~\ref{alg:func-begin}--\ref{alg:func-end} builds the circuit for
each AND gate $x_i$ as illustrated in Fig.~\ref{fig:and-step}.

\SetKwProg{Fn}{function}{ is}{end}

\begin{algorithm}
  \KwIn{Logic network with gates $x_{n+1}, \dots, x_{n+r}$}
  \KwOut{Quantum circuit for $U_f$}

  \Fn{compute}{
  \For{$i = n + 1, \dots, n+r$ where $\circ_i = \land$}{%
    set $p \gets p(i)$, $q \gets q(i)$, $j \gets j(i)$, $k \gets k(i)$\; \label{alg:func-begin}
    set $L_1 \gets \ltfi(x_j)$, $L_2 \gets \ltfi(x_k)$\;
    \If{$L_1 \subseteq L_2$}{%
      swap $L_1 \leftrightarrow L_2$ and $p \leftrightarrow q$\;
    }
    let $t_1$ be some element in $L_1 \setminus L_2$\;
    let $t_2$ be some element in $L_2$\;
    CNOT$(x, t_1)$ for all $x \in L_1 \setminus \{t_1\}$\;
    CNOT$(x, t_2)$ for all $x \in L_2 \setminus \{t_2\}$\;
    \lIf{$p$}{NOT$(t_1)$}
    \lIf{$q$}{NOT$(t_2)$}
    TOFFOLI$(t_1, t_2, x_i)$\;
    \lIf{$p$}{NOT$(t_2)$}
    \lIf{$q$}{NOT$(t_1)$}
    CNOT$(x, t_2)$ for all $x \in L_2 \setminus \{t_2\}$\;
    CNOT$(x, t_1)$ for all $x \in L_1 \setminus \{t_1\}$\; \label{alg:func-end}
  }}
  compute\; \label{alg:main-begin}
  CNOT$(x_{n+r}, y)$\;
  \lIf{$p = 0$}{NOT$(y)$}
  compute$^\dagger$\; \label{alg:main-end}
  \caption{Heuristic compilation algorithm.}
  \label{alg:heuristic}
\end{algorithm}

Note that we assume that $L_1 \neq L_2$.  If this is not the case, it
means that the functions computed by fan-in to the AND gate are equal,
making the AND gate redundant.  Also, note that the intersection of
$L_1$ and $L_2$ may not be empty.  Since we want to compute the value
of $L_1$ in-place on some signal $t_1 \in L_1$, we must ensure that
$L_1 \not\subseteq L_2$.  If the latter condition applies, it is sufficient to swap
$L_1$ and $L_2$.

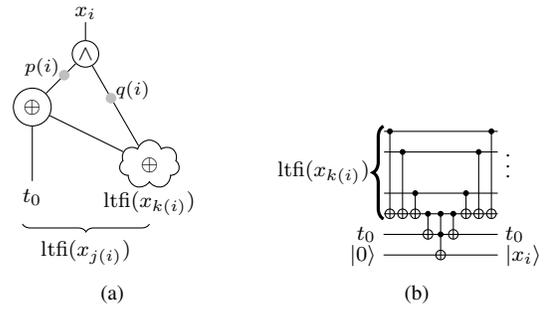
\begin{figure}[t!]
  \centering
  \subfloat[]{%
    \begin{tikzpicture}[font=\footnotesize]
      \node[inner sep=1pt,draw,circle] (xi) {$\land$};
      \draw (xi) -- ++(up:12pt) node[above,inner sep=1pt] {$x_i$};
      \node[draw,circle, inner sep = 2pt] (cloud1) at ($(xi)+(225:1cm)$) {$\oplus$};
      \node[draw,cloud,cloud puffs=7,inner sep=2pt,minimum width=.8cm] (cloud2) at ($(xi)+(300:1.7cm)$) {$\oplus$};
      \draw (xi) -- (cloud1) (xi) -- (cloud2) (cloud1) -- (cloud2);
      \draw (cloud1) -- ++(down:28pt) coordinate (so) node [below, inner sep = 3pt](bo){$t_0$};
      \node[below,inner sep=3pt](ltk) at (cloud2.south) {$\ltfi(x_{k(i)})$};
      \draw[decorate,decoration={brace,amplitude=3pt}] 
    (ltk.south) -- ++ (left:48pt) coordinate (t_k_opt_impl_unten); 
      \node[below= 3pt, xshift= 24pt,inner sep=3pt] at (t_k_opt_impl_unten.south) {$\ltfi(x_{j(i)})$};
      \fill[gray!50!white] ($(xi)!.4!(cloud1)$) circle (2pt) node[left,yshift=3pt,inner sep=1.5pt,black,font=\scriptsize] {$p(i)$};
      \fill[gray!50!white] ($(xi)!.4!(cloud2)$) circle (2pt) node[right,yshift=3pt,inner sep=1.5pt,black,font=\scriptsize] {$q(i)$};
    \end{tikzpicture}
    }
  \hfil
  \subfloat[]{%
    \footnotesize
\begin{tikzpicture}[scale=0.600000,x=1pt,y=1pt]
\filldraw[color=white] (0.000000, -6.500000) rectangle (72.000000, 84.500000);
\draw[color=black] (0.000000,78.000000) -- (72.000000,78.000000);
\draw[color=black] (0.000000,65.000000) -- (72.000000,65.000000);
\draw[color=black] (0.000000,52.000000) node[left] {${}$};
\draw[color=black] (0.000000,39.000000) -- (72.000000,39.000000);
\draw[color=black] (0.000000,26.000000) -- (72.000000,26.000000);
\filldraw[color=white,fill=white] (0.000000,22.750000) rectangle (-4.000000,81.250000);
\draw[decorate,decoration={brace,amplitude = 4.000000pt},very thick] (0.000000,22.750000) -- (0.000000,81.250000);
\draw[color=black] (-4.000000,52.000000) node[left] {${\operatorname{ltfi}(x_{k(i)})}$};
\draw[color=black] (0.000000,13.000000) -- (72.000000,13.000000);
\draw[color=black] (0.000000,13.000000) node[left] {$t_0$};
\draw[color=black] (0.000000,0.000000) -- (72.000000,0.000000);
\draw[color=black] (0.000000,0.000000) node[left] {$|0\rangle$};
\draw (4.000000,78.000000) -- (4.000000,26.000000);
\filldraw (4.000000, 78.000000) circle(1.500000pt);
\begin{scope}
\draw[fill=white] (4.000000, 26.000000) circle(3.000000pt);
\clip (4.000000, 26.000000) circle(3.000000pt);
\draw (1.000000, 26.000000) -- (7.000000, 26.000000);
\draw (4.000000, 23.000000) -- (4.000000, 29.000000);
\end{scope}
\draw (12.000000,65.000000) -- (12.000000,26.000000);
\filldraw (12.000000, 65.000000) circle(1.500000pt);
\begin{scope}
\draw[fill=white] (12.000000, 26.000000) circle(3.000000pt);
\clip (12.000000, 26.000000) circle(3.000000pt);
\draw (9.000000, 26.000000) -- (15.000000, 26.000000);
\draw (12.000000, 23.000000) -- (12.000000, 29.000000);
\end{scope}
\draw (20.000000,39.000000) -- (20.000000,26.000000);
\filldraw (20.000000, 39.000000) circle(1.500000pt);
\begin{scope}
\draw[fill=white] (20.000000, 26.000000) circle(3.000000pt);
\clip (20.000000, 26.000000) circle(3.000000pt);
\draw (17.000000, 26.000000) -- (23.000000, 26.000000);
\draw (20.000000, 23.000000) -- (20.000000, 29.000000);
\end{scope}
\draw (28.000000,26.000000) -- (28.000000,13.000000);
\filldraw (28.000000, 26.000000) circle(1.500000pt);
\begin{scope}
\draw[fill=white] (28.000000, 13.000000) circle(3.000000pt);
\clip (28.000000, 13.000000) circle(3.000000pt);
\draw (25.000000, 13.000000) -- (31.000000, 13.000000);
\draw (28.000000, 10.000000) -- (28.000000, 16.000000);
\end{scope}
\draw (36.000000,26.000000) -- (36.000000,0.000000);
\filldraw (36.000000, 26.000000) circle(1.500000pt);
\filldraw (36.000000, 13.000000) circle(1.500000pt);
\begin{scope}
\draw[fill=white] (36.000000, 0.000000) circle(3.000000pt);
\clip (36.000000, 0.000000) circle(3.000000pt);
\draw (33.000000, 0.000000) -- (39.000000, 0.000000);
\draw (36.000000, -3.000000) -- (36.000000, 3.000000);
\end{scope}
\draw (44.000000,26.000000) -- (44.000000,13.000000);
\filldraw (44.000000, 26.000000) circle(1.500000pt);
\begin{scope}
\draw[fill=white] (44.000000, 13.000000) circle(3.000000pt);
\clip (44.000000, 13.000000) circle(3.000000pt);
\draw (41.000000, 13.000000) -- (47.000000, 13.000000);
\draw (44.000000, 10.000000) -- (44.000000, 16.000000);
\end{scope}
\draw (52.000000,39.000000) -- (52.000000,26.000000);
\filldraw (52.000000, 39.000000) circle(1.500000pt);
\begin{scope}
\draw[fill=white] (52.000000, 26.000000) circle(3.000000pt);
\clip (52.000000, 26.000000) circle(3.000000pt);
\draw (49.000000, 26.000000) -- (55.000000, 26.000000);
\draw (52.000000, 23.000000) -- (52.000000, 29.000000);
\end{scope}
\draw (60.000000,65.000000) -- (60.000000,26.000000);
\filldraw (60.000000, 65.000000) circle(1.500000pt);
\begin{scope}
\draw[fill=white] (60.000000, 26.000000) circle(3.000000pt);
\clip (60.000000, 26.000000) circle(3.000000pt);
\draw (57.000000, 26.000000) -- (63.000000, 26.000000);
\draw (60.000000, 23.000000) -- (60.000000, 29.000000);
\end{scope}
\draw (68.000000,78.000000) -- (68.000000,26.000000);
\filldraw (68.000000, 78.000000) circle(1.500000pt);
\begin{scope}
\draw[fill=white] (68.000000, 26.000000) circle(3.000000pt);
\clip (68.000000, 26.000000) circle(3.000000pt);
\draw (65.000000, 26.000000) -- (71.000000, 26.000000);
\draw (68.000000, 23.000000) -- (68.000000, 29.000000);
\end{scope}
\draw[color=black] (72.000000,52.000000) node[anchor=mid west] {$\vdots$};
\draw[color=black] (72.000000,13.000000) node[right] {$t_0$};
\draw[color=black] (72.000000,0.000000) node[right] {$|x_i\rangle$};
\end{tikzpicture}
  }
  \caption{Example in which one transitive fan-in in included in the other. The computed values can be reused.}
  \label{fig:incl-step}
\end{figure}

In addition, when $L_2 \subseteq L_1 $, the value computed by $L_2$ could be reused to compute $L_1$. 
This is achieved by modifying the elements in $L_1$ such that $L_1 = (L_1 \setminus L_2) \cup \{x_k\}$.
An example is shown in Fig.~\ref{fig:incl-step}. In this case $\ltfi(x_{j})$ includes $\ltfi(x_{k})$ and  $\ltfi(x_{j}) \setminus \ltfi(x_{k}) = \{t_0\}$. 
In general, when this optimization applies, it allows us to save $2 \cdot |\ltfi(x_{k})|$ CNOT operations.

\section{Pebble Strategies}
\label{sec:pebbling}
In this section, we describe a dedicated SAT-based reversible pebbling
strategy for logic networks over the basis $\{\land, \oplus, \neg\}$.
A SAT-based reversible pebbling strategy allows us to reduce the
number of qubits by trading off $T$-count.  Compared to existing
reversible pebbling strategies (see, e.g., \cite{PRS17,Meuli19}), we
enforce that all XOR operations are performed in-place.

\begin{figure}[t]
  \centering
  \subfloat[for AND gates $x_i$]{%
    \begin{tikzpicture}[font=\footnotesize]
      \node[inner sep=1pt,draw,circle] (xi) {$\phantom{\land}$};
      \node[inner sep=1pt,draw,circle] (xj) at ($(xi)+(235:.7cm)$) {$\phantom{\land}$};
      \node[inner sep=1pt,draw,circle] (xk) at ($(xi)+(305:.7cm)$) {$\phantom{\land}$};
      \node[above right] at (xi) {$x_i$};
      \node[above left] at (xj) {$x_{j(i)}$};
      \node[above right] at (xk) {$x_{k(i)}$};

      \draw (xi) -- (xj);
      \draw (xi) -- (xk);
      \fill (xj) circle (2pt);
      \fill (xk) circle (2pt);

      \node[inner sep=1pt,draw,circle] (xi2) at (3.2,0) {$\phantom{\land}$};
      \node[inner sep=1pt,draw,circle] (xj2) at ($(xi2)+(235:.7cm)$) {$\phantom{\land}$};
      \node[inner sep=1pt,draw,circle] (xk2) at ($(xi2)+(305:.7cm)$) {$\phantom{\land}$};
      \node[above right] at (xi2) {$x_i$};
      \node[above left] at (xj2) {$x_{j(i)}$};
      \node[above right] at (xk2) {$x_{k(i)}$};

      \draw (xi2) -- (xj2);
      \draw (xi2) -- (xk2);
      \fill (xi2) circle (2pt);
      \fill (xj2) circle (2pt);
      \fill (xk2) circle (2pt);

      \node at (1.6,-.25) {$\longleftrightarrow$};
    \end{tikzpicture}
  }
  \hfil
  \subfloat[for XOR gates $x_i$]{%
    \begin{tikzpicture}[font=\footnotesize]
      \node[inner sep=1pt,draw,circle] (xi) {$\phantom{\land}$};
      \node[inner sep=1pt,draw,circle] (xj) at ($(xi)+(235:.7cm)$) {$\phantom{\land}$};
      \node[inner sep=1pt,draw,circle] (xk) at ($(xi)+(305:.7cm)$) {$\phantom{\land}$};
      \node[above right] at (xi) {$x_i$};
      \node[above left] at (xj) {$x_{j(i)}$};
      \node[above right] at (xk) {$x_{k(i)}$};

      \draw (xi) -- (xj);
      \draw (xi) -- (xk);
      \fill (xj) circle (2pt);
      \fill (xk) circle (2pt);

      \node[inner sep=1pt,draw,circle] (xi2) at (3.2,0) {$\phantom{\land}$};
      \node[inner sep=1pt,draw,circle] (xj2) at ($(xi2)+(235:.7cm)$) {$\phantom{\land}$};
      \node[inner sep=1pt,draw,circle] (xk2) at ($(xi2)+(305:.7cm)$) {$\phantom{\land}$};
      \node[above right] at (xi2) {$x_i$};
      \node[above left] at (xj2) {$x_{j(i)}$};
      \node[above right] at (xk2) {$x_{k(i)}$};

      \draw (xi2) -- (xj2);
      \draw (xi2) -- (xk2);
      \fill (xi2) circle (2pt);
      \fill (xj2) circle (2pt);

      \node[inner sep=1pt,draw,circle] (xi3) at (-3.2,0) {$\phantom{\land}$};
      \node[inner sep=1pt,draw,circle] (xj3) at ($(xi3)+(235:.7cm)$) {$\phantom{\land}$};
      \node[inner sep=1pt,draw,circle] (xk3) at ($(xi3)+(305:.7cm)$) {$\phantom{\land}$};
      \node[above right] at (xi3) {$x_i$};
      \node[above left] at (xj3) {$x_{j(i)}$};
      \node[above right] at (xk3) {$x_{k(i)}$};

      \draw (xi3) -- (xj3);
      \draw (xi3) -- (xk3);
      \fill (xi3) circle (2pt);
      \fill (xk3) circle (2pt);

      \node at (-1.6,-.25) {$\longleftrightarrow$};
      \node at (1.6,-.25) {$\longleftrightarrow$};
    \end{tikzpicture}
  }
  \caption{Pebble moves.}
  \label{fig:pebble-moves}
\end{figure}
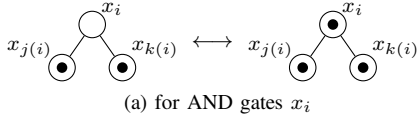
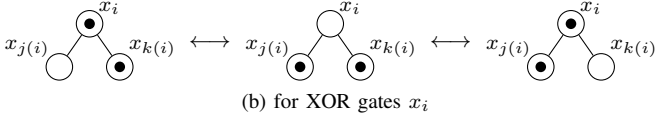

The reversible pebble game is played on a logic network.  Each node in
a logic network can be assigned a pebble or not.  We say that a node
is pebbled, if it is assigned a pebble.  At the beginning of the
pebble game, at time step $s = 0$, all the primary inputs of the logic
network are pebbled and all the gates are not.  At each step a pebble
can be put or removed from an AND gate, if both children are pebbled.
For an XOR gate, the output can be pebbled, if both inputs are
pebbled, but afterwards one of the input pebbles must be removed.
This indicates that the result of the XOR computation (using a
$\text{\small CNOT}$ gate) has been computed into that input.  All
moves can be performed in a bidirectional way.
Fig.~\ref{fig:pebble-moves} provides a summary.  We allow multiple
moves in a single step.  The game is won when all output gates and all
input nodes are pebbled, but all other nodes are not.  Each pebble
corresponds to a qubit that currently stores the computation result.
Therefore, we are interested in the resource-constrained reversible
pebble game in which at each step one can only use at most $L$
pebbles.

We describe the SAT formula for the case in which the logic network
drives a single output with gate $x_{n+r}$.  The SAT formula has
variables $x_i^{(s)}$ for each node $x_i$ and step $0 \le s \le S$ in
the network.  The value of $x_i^{(s)}$ encodes whether node $x_i$ is
pebbled at step $s$.  Initially, all inputs are pebbled while all
other nodes are not.  At the last step $S$, the primary inputs and the
output node must be the only nodes that are pebbled.  These are enforced
by the unit clauses
\begin{equation}
  \label{eq:cnf-init}
  x_i^{(0)} \oplus [i > n]
\end{equation}
and
\begin{equation}
  \label{eq:cnf-final}
  x_{i}^{(S)} \oplus [n < i < n + r]
\end{equation}
for all $1 \le i \le n + r$, respectively, where $[\cdot]$ denotes the
Iverson bracket.  For all gates $x_i$ such that $\circ_i = \land$, the
constraint
\begin{equation}
  \label{eq:cnf-and}
  (x_i^{(s)} \oplus x_i^{(s+1)}) \rightarrow
  (x_{j(i)}^{(s)} \land x_{j(i)}^{(s+1)} \land x_{k(i)}^{(s)} \land x_{k(i)}^{(s+1)})
\end{equation}
ensures that the pebble of AND gate $x_i$ can only change from time
$s$ to $s+1$, if all two children are pebbled in both time steps.

For all gates $x_i$ such that $\circ_i = \oplus$, the constraints
\begin{equation}
  \label{eq:cnf-xor1}
  (\bar x_i^{(s)} \land x_i^{(s+1)}) \rightarrow
  (x_{j(i)}^{(s)} \land x_{k(i)}^{(s)} \land (x_{j(i)}^{(s+1)} \oplus x_{k(i)}^{(s+1)}))
\end{equation}
and
\begin{equation}
  \label{eq:cnf-xor1}
  (x_i^{(s)} \land \bar x_i^{(s+1)}) \rightarrow
  ((x_{j(i)}^{(s)} \oplus x_{k(i)}^{(s)}) \land x_{j(i)}^{(s+1)} \land x_{k(i)}^{(s+1)})
\end{equation}
ensure the reversible pebble game semantics for XOR gates in
Fig.~\ref{fig:pebble-moves}(b).

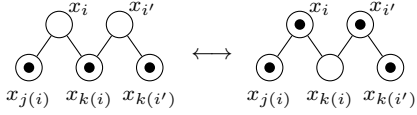
\begin{figure}[t]
  \centering
  \begin{tikzpicture}[font=\footnotesize]
    \node[inner sep=1pt,draw,circle] (xi) {$\phantom{\land}$};
    \node[inner sep=1pt,draw,circle] (xj) at ($(xi)+(235:.7cm)$) {$\phantom{\land}$};
    \node[inner sep=1pt,draw,circle] (xk) at ($(xi)+(305:.7cm)$) {$\phantom{\land}$};
    \node[inner sep=1pt,draw,circle] (xm) at ($(xk)+(55:.7cm)$) {$\phantom{\land}$};
    \node[inner sep=1pt,draw,circle] (xn) at ($(xm)+(305:.7cm)$) {$\phantom{\land}$};
    \node[above right] at (xi) {$x_i$};
    \node[above right] at (xm) {$x_{i'}$};
    \node[below] at (xj.south) {$x_{j(i)}$};
    \node[below] at (xk.south) {$x_{k(i)}$};
    \node[below] at (xn.south) {$x_{k(i')}$};

    \draw (xi) -- (xj);
    \draw (xi) -- (xk);
    \draw (xm) -- (xk);
    \draw (xm) -- (xn);
    \fill (xj) circle (2pt);
    \fill (xk) circle (2pt);
    \fill (xn) circle (2pt);

    \node[inner sep=1pt,draw,circle] (xi2) at (3.2,0) {$\phantom{\land}$};
    \node[inner sep=1pt,draw,circle] (xj2) at ($(xi2)+(235:.7cm)$) {$\phantom{\land}$};
    \node[inner sep=1pt,draw,circle] (xk2) at ($(xi2)+(305:.7cm)$) {$\phantom{\land}$};
    \node[inner sep=1pt,draw,circle] (xm2) at ($(xk2)+(55:.7cm)$) {$\phantom{\land}$};
    \node[inner sep=1pt,draw,circle] (xn2) at ($(xm2)+(305:.7cm)$) {$\phantom{\land}$};
    \node[above right] at (xi2) {$x_i$};
    \node[above right] at (xm2) {$x_{i'}$};
    \node[below] at (xj2.south) {$x_{j(i)}$};
    \node[below] at (xk2.south) {$x_{k(i)}$};
    \node[below] at (xn2.south) {$x_{k(i')}$};

    \draw (xi2) -- (xj2);
    \draw (xi2) -- (xk2);
    \draw (xm2) -- (xk2);
    \draw (xm2) -- (xn2);
    \fill (xi2) circle (2pt);
    \fill (xn2) circle (2pt);
    \fill (xj2) circle (2pt);
    \fill (xm2) circle (2pt);

    \node at (current bounding box.center) {$\longleftrightarrow$};
  \end{tikzpicture}
  \caption{Illegal XOR move for pair of XOR gates $x_i$ and $x_{i'}$.}
  \label{fig:xor-conflict-case}
\end{figure}

Since we allow multiple moves at a single step, the current constraits
so far allow illegal moves such as the one illustrated in
Fig.~\ref{fig:xor-conflict-case}.  Here, $x_i$ and $x_{i'}$ are two
XOR gates with a shared child $x_{k(i)} = x_{j(i')}$.  We rule out
such cases with explicit blocking constraints
\begin{multline}
  \label{eq:illegal-xor}
  \overline{\left(\bar x_i^{(s)} \land \bar x_{i'}^{(s)} \land x_{c}^{(s)} \land x_i^{(s+1)} \land x_{i'}^{(s+1)} \land \bar x_{c}^{(s+1)}\right)} \\
  \overline{\left(x_i^{(s)} \land x_{i'}^{(s)} \land \bar x_{c}^{(s)} \land \bar x_i^{(s+1)} \land \bar x_{i'}^{(s+1)} \land x_{c}^{(s+1)}\right)},
\end{multline}
where $\circ_i = \circ_{i'} = \oplus$ and
$\{c\} = \{j(i), k(i)\} \cap \{j(i'), k(i')\}$ is the common child of
gate $x_i$ and $x_{i'}$.

All constraints described
in~\eqref{eq:cnf-and}--\eqref{eq:illegal-xor} can easily be expressed
as a CNF (conjunctive normal form).  The final clause
\begin{equation}
  \label{eq:card}
  \sum_{i=1}^{n+r} x_i^{(s)} \le L
\end{equation}
for each $1 \le s \le S$ to restrict the number of used pebbles at
each step $s$ is translated into a CNF using cardinality constraints
(see, e.g.,~\cite{Knuth4f6}).  We employ a bounded-model checking
style procedure using incremental SAT solving where we increase the
maximum number of steps $S$ until a solution is found, or terminate
once a given resource limit is reached.

\section{Experimental Results}
In our experiments, we compile quantum oracles for functions represented by XAGs, targeting quantum circuits over the Clifford+$T$ gate set. Our constructive algorithm and the SAT-based pebbling technique are implemented in the open-source C++ library for quantum compilation \emph{caterpillar}\footnote{https://github.com/gmeuli/caterpillar}, that 
empowers \emph{RevKit 3.1}\footnote{https://github.com/msoeken/revkit}.

Our synthesis approach is capable of generating circuits with exactly $4\cdot \tilde c$ $T$ gates, where $\tilde c$ is the number of AND nodes in the network.
In cases in which the XAG implements the function with the minimum possible number of AND nodes ($\tilde c = c_{\wedge}(f)$), our method returns a circuit with the minimum $T$-count. As indicated in Table~\ref{tab_epfl} and~\ref{table_crypt} by `$\ast$', all the adders (whose XAGs are proven to have minimum multiplicative complexity~\cite{BP08}) are synthesized into circuits with the same $T$-count of the best-known manually designed circuits~\cite{Gidney18}.

We compare against state-of-the-art methods that are also implemented in \emph{RevKit 3.1}, synthesizing combinatorial circuits from the EPFL benchmarks,\footnote{https://lsi.epfl.ch/benchmarks} both arithmetic and random control.
In the results, we ignore Clifford gates as it is customary in fault-tolerant computing~\cite{AMMR13}. However, assuming that a $T$ gate costs $50\times$ as much as a Clifford gate, we would still have an overall gain compared to the state-of-the-art.

\subsection{Comparing to best-fit LUT-based synthesis}
The first comparison is performed gainst the LUT-based hierarchical synthesis method \emph{best-fit LHRS}~\cite{Meuli18}.

Look-up table (LUT) mapping is a logic network decomposition technique widely used for logic circuit optimization~\cite{Mishchenko06}.
A $k$-LUT mapping decomposes a logic network into $k$-feasible LUTs, i.e., single-output subnetworks with maximum $k$ inputs. 
Hierarchical synthesis methods for quantum circuits use the LUT-mapping to decompose the design such that less scalable methods, e.g., ESOP-based synthesis~\cite{fazel07}, can be applied. This decomposition step defines the final number of qubits used by the circuit, that can be controlled by the parameter $k$. In fact, a large $k$ will give fewer LUTs with more variables and, as a consequence, less qubits will be needed to store intermediate results. This procedure is typical of any LUT-based hierarchical method. In our experiment we set $k$ to 16.  The \emph{best-fit} method uses an additional LUT mapping step for the synthesis of each sub-network. Thus allowing a reduced number of $T$ gates with respect to similar techniques.

Our results show that the overhead in $T$-count that this method requires to reduce the number of qubits is too large, if compared with our technique, as shown in Table~\ref{tab_epfl}. This is true, especially considering that among the hierarchical methods, \emph{best-fit} is the one returning the lowest number of $T$ gates.
Our results have in average about $20.4 \times$ smaller $T$-count, while the number of qubits is doubled. In the context of fault tolerant quantum computing, where the number of $T$ gates is the predominant cost metric, our method outperforms the \emph{best-fit} LUT-based method.

With our approach, the circuits count more Clifford gates. Nevertheless, if we consider the Clifford gates, together with the $T$ gates, and accounting a 50:1 cost ratio, we still get that the LUT-based results are worst, in average, with respect to our approach.
In addition, our current heuristic compilation technique is not exploiting shared logic among the CNOT gates, besides the extreme case discussed in Section~\ref{comp. alg.}. A more careful analysis can lead to further CNOT gate count reduction.

\subsection{Comparing to hierarchical synthesis with Bennett clean-up}
The second comparison we present is with respect to a state-of-the-art hierarchical method that uncomputes ancillae using the Bennett strategy~\cite{Bennett89}: nodes are synthesized in a bottom-up order, and uncomputation is performed in a top-down order. To make our comparison fair, we apply this method using XAG networks as inputs (a different network choice, e.g., and-inverter graphs, would lead to higher $T$-count). In addition, we modify this technique to also exploit the 4 $T$ gates quantum AND implementation (see Section~\ref{quant. circ.}).
This hierarchihcal method synthesizes a Toffoli gate for each AND node, two {\small CNOT} gates for each XOR node, and an $X$ gate for each inversion, during computation and uncomputation. It results in circuits with the same minimum number of $T$ gates achieved by our method. This minimum number is proportional to $\tilde{c}$---the number of AND nodes in the specification network. 
As shown in Table~\ref{tab_epfl}, we achieve $70\%$ fewer qubits on average. This is due to the strategy of computing XOR-blocks \emph{in-place}, without using any ancilla as shown in Fig.~\ref{fig:and-step}. 

\begin{table*}[t]
\centering 
\caption{Comparisons}
\scriptsize
\begin{tabularx}{\textwidth}{Xrrrrrrrrrrrrrrrr}
\toprule
     &       &        &          &          &  \multicolumn{4}{c}{\textbf{Best-fit lhrs~\cite{Meuli18}}}    &     \multicolumn{4}{c}{\textbf{Bennett}}   & \multicolumn{4}{c}{\textbf{Proposed}}\\
     \cmidrule(lr){6-9}
     \cmidrule(lr){10-13}
     \cmidrule(lr){14-17}
  &  I  &  O  &  XOR  &  AND  &  qubits  & T  &  Clifford   &   t[s]  &  qubits  &  T  &  Clifford  &    t[s]  &qubits  &   T  &  Clifford  &   t[s] \\
adder    &  256  &  129  &  549  &  128  &  448  &  14411  &  319  &  0.0  &  933  &  512  &  1938  &  0.0  &  385  &  512$\ast$  &  1778  &  0.0  \\   
arbiter    &  256  &  129  &  0  &  1181  &  694  &  134492  &  1297  &  0.1  &  1437  &  4724  &  1  &  0.1  &  1437  &  4724  &  1  &  0.1  \\    
bar    &  135  &  128  &  1728  &  832  &  863  &  44800  &  1328  &  0.1  &  2695  &  3328  &  6656  &  0.4  &  1032  &  3328  &  23552  &  0.3  \\  
cavlc    &  10  &  11  &  197  &  494  &  123  &  59650  &  430  &  0.0  &  701  &  1976  &  788  &  0.0  &  504  &  1976  &  600  &  0.1  \\    
ctrl    &  7  &  26  &  8  &  85  &  36  &  4102  &  21  &  0.0  &  101  &  340  &  41  &  0.0  &  93  &  340  &  21  &  0.0  \\  
dec    &  8  &  256  &  0  &  341  &  293  &  30061  &  58  &  0.0  &  349  &  1364  &  0  &  0.0  &  349  &  1364  &  0  &  0.0  \\  
div    &  128  &  128  &  8994  &  6060  &  4188  &  248739  &  10044  &  1.1  &  15182  &  24240  &  35891  &  13.7  &  6188  &  24240  &  260301  &  14.2  \\ 
i2c    &  147  &  142  &  502  &  623  &  377  &  65674  &  604  &  0.0  &  1273  &  2492  &  2041  &  0.1  &  771  &  2492  &  1247  &  0.1  \\  
int2float    &  11  &  7  &  101  &  100  &  44  &  8645  &  104  &  0.0  &  212  &  400  &  405  &  0.0  &  111  &  400  &  345  &  0.0  \\  
log2    &  32  &  32  &  9371  &  19436  &  8192  &  2177089  &  28176  &  4.4  &  28839  &  77744  &  37437  &  46.5  &  19469  &  77744  &  965225  &  65.3  \\ 
max    &  512  &  130  &  1479  &  931  &  1076  &  82287  &  2265  &  0.1  &  2922  &  3724  &  5658  &  0.4  &  1444  &  3724  &  9060  &  0.4  \\    
mem\_ctrl    &  1204  &  1231  &  4168  &  5113  &  3397  &  498240  &  5206  &  0.3  &  10486  &  20452  &  15899  &  4.1  &  6319  &  20452  &  12077  &  5.6  \\    
multiplier    &  128  &  128  &  8614  &  11940  &  5294  &  839571  &  21330  &  1.8  &  20682  &  47760  &  34269  &  24.4  &  12069  &  47760  &  869535  &  39.5  \\    
priority    &  128  &  8  &  158  &  327  &  256  &  65563  &  571  &  0.0  &  613  &  1308  &  633  &  0.0  &  455  &  1308  &  467  &  0.0  \\   
router    &  60  &  30  &  0  &  96  &  90  &  7930  &  93  &  0.0  &  157  &  384  &  27  &  0.0  &  157  &  384  &  27  &  0.0  \\   
sin    &  24  &  25  &  1770  &  4075  &  1531  &  392926  &  5881  &  0.3  &  5869  &  16300  &  7046  &  2.0  &  4099  &  16300  &  62330  &  3.5  \\    
sqrt    &  128  &  64  &  9640  &  6244  &  4297  &  375639  &  19207  &  1.2  &  16012  &  24976  &  38609  &  14.5  &  6372  &  24976  &  207921  &  15.1  \\ 
square    &  64  &  128  &  8084  &  5181  &  3967  &  262544  &  8905  &  1.0  &  13330  &  20724  &  32154  &  10.0  &  5246  &  20724  &  247002  &  8.6  \\    
voter    &  1001  &  1  &  6066  &  5651  &  2640  &  274431  &  7084  &  0.3  &  12718  &  22604  &  24265  &  7.9  &  6652  &  22604  &  168081  &  9.5\\

\midrule
\multicolumn{5}{l}{\textbf{Normalized geometric mean} } & \textbf{0.5}  &  \textbf{20.4 } &    &    &  \textbf{1.7 } & \textbf{ 1}  &    &    &  \textbf{1}  &  \textbf{1  }&    &  \\
\bottomrule
\multicolumn{17}{l}{ }\\
\multicolumn{17}{l}{ $\ast $ matches best-known $T$-count~\cite{Gidney18}.}
\end{tabularx}

\label{tab_epfl}
\end{table*}

\begin{table}[t]
\centering 
\caption{Benchmarks from cryptography}
\def\tabcolsep{3pt}
\scriptsize
\begin{tabularx}{\columnwidth}{Xrrrrrrrr}
\toprule
  &  I  &  O  &  XOR  &  AND  &  qubits  &  Clifford  &  T  &  t[s]  \\
  \midrule
bm\_10  &  20  &  19  &  102  &  52  &  89  &  770  &  208  &  0.0 \\  
bm\_11  &  22  &  21  &  108  &  78  &  119  &  716  &  312  &  0.0  \\
bm\_12  &  24  &  23  &  126  &  81  &  120  &  908  &  324  &  0.0  \\
bm\_15  &  30  &  29  &  195  &  117  &  174  &  1776  &  468  &  0.0 \\ 
bm\_20  &  40  &  39  &  314  &  208  &  279  &  3154  &  832  &  0.0  \\
bm\_30  &  60  &  59  &  687  &  351  &  452  &  8756  &  1404  &  0.1 \\
bm\_40  &  80  &  79  &  1079  &  624  &  759  &  15618  &  2496  &  0.2  \\
bm\_50  &  100  &  99  &  1847  &  676  &  855  &  32354  &  2704  &  0.3  \\
bm\_60  &  120  &  119  &  2253  &  1053  &  1262  &  41324  &  4212  &  0.5 \\  
bm\_70  &  140  &  139  &  2985  &  1432  &  1643  &  54036  &  5728  &  0.8  \\
bm\_80  &  160  &  159  &  3494  &  1872  &  2151  &  73570  &  7488  &  1.4  \\
bm\_90  &  180  &  179  &  4561  &  1989  &  2318  &  105578  &  7956  &  1.6  \\ 
bm\_100  &  200  &  199  &  5143  &  2704  &  3063  &  129810  &  10816  &  2.7  \\ 
mcustom  &  16  &  8  &  79  &  27  &  51  &  424  &  108  &  0.0  \\  
mx6x31  &  12  &  6  &  30  &  27  &  45  &  132  &  108  &  0.0  \\ 
mx7x41  &  14  &  7  &  44  &  40  &  61  &  156  &  160  &  0.0  \\
mx7x41  &  14  &  7  &  45  &  40  &  61  &  168  &  160  &  0.0  \\  
s16  &  17  &  16  &  333  &  113  &  146  &  7049  &  452  &  0.0  \\ 
s8& 8& 8& 83& 32& 48& 1408& 128& 0.0\\
x8x4x31  &  16  &  8  &  69  &  48  &  72  &  370  &  192  &  0.0  \\
\midrule
adder\_32bit  &  64  &  33  &  150  &  32  &  97  &  556  &  128$\ast$  &  0.0  \\ 
adder\_64bit  &  128  &  65  &  284  &  64  &  193  &  1132  &  256$\ast$  &  0.0  \\  
AES-expanded  &  1536  &  128  &  20325  &  5440  &  6979  &  1583232  &  21760  &  14.2  \\
AES-non-expanded  &  256  &  128  &  25124  &  6800  &  7059  &  3009160  &  27200  &  22.2  \\
comp\_32bit\_sign\_LT  &  64  &  1  &  116  &  108  &  172  &  312  &  432  &  0.0  \\  
comp\_32bit\_sign\_LTEQ  &  64  &  1  &  89  &  114  &  178  &  265  &  456  &  0.0  \\  
DES-expanded  &  832  &  64  &  11263  &  15126  &  15959  &  256441  &  60504  &  69.4  \\
DES-non-expanded  &  128  &  64  &  11105  &  15093  &  15222  &  246564  &  60372  &  68.7  \\
mult-32x32 & 64 & 64  & 2473 & 4107 & 4172 & 54498 &16428 & 3.0 \\
\bottomrule
\multicolumn{9}{l}{ }\\
\multicolumn{9}{l}{ $\ast $ matches best-known $T$-count~\cite{Gidney18}.}
\end{tabularx}
\label{table_crypt}
\end{table}

\subsection{Minimized circuits with cryptography applications}
With our compilation technique, we synthesize quantum oracles for Boolean functions used in a wide range of applications, from encryption to digital signatures, from hashing to error correction codes. The best-known version of these circuits, in terms of multiplicative complexity and depth, have been collected by the \emph{Computer Security Resource Center} (CSRC) at the \emph{National Institute of Standards and Technology} (NIST).
Our results are shown in Table~\ref{table_crypt}. We synthesize: (i) finite field multiplication in $GF(2^6)$ using irreducible polynomial $x^6 + x^3 + 1$ (\emph{mx6x31}), multiplication in $GF(2^7)$ using irreducible polynomial $x^7 + x^4 + 1$ (\emph{mx7x41}) and using $x^7 + x^3 + 1$ (\emph{mx7x41}); (ii) binary multiplication with different input size $n$ (\emph{bm\_n}); (iii) a 16-bit and a 8-bit S-box (\emph{s16}, \emph{s8}); (iv) finite field multiplication in $GF(2^8)$ using the AES polynomial $x^8 + x^4 + x^3 + x + 1$ (\emph{x8x4x31}). The benchmark circuits are available online.\footnote{http://cs-www.cs.yale.edu/homes/peralta/CircuitStuff/CMT.html}

In addition, we evaluated our method on a set of circuits used in the context of \emph{Multi-Party Computation} MPC and \emph{Fully Homomorphic Encryption} FHE, optimized for the number of AND gates using the technique proposed in~\cite{testa19}. Oracles for these functions can be used to evaluate the cost of a Grover's attack over the relative encryption scheme. In fact, it has been shown how Grover's algorithm can be used as the quantum version of a key attack, if the quantum circuit for the encryption function is known~\cite{grassl16}. From the benchmarks available online\footnote{https://homes.esat.kuleuven.be/\char`\~nsmart/MPC/} we synthesize: (i) block ciphers AES and DES in their expanded and non expanded variant, the latter meaning that the input key is assumed non-expanded; (ii) arithmetic functions such as adders, multipliers, and comparators.

\subsection{Evaluating the pebble strategies for the S-box}
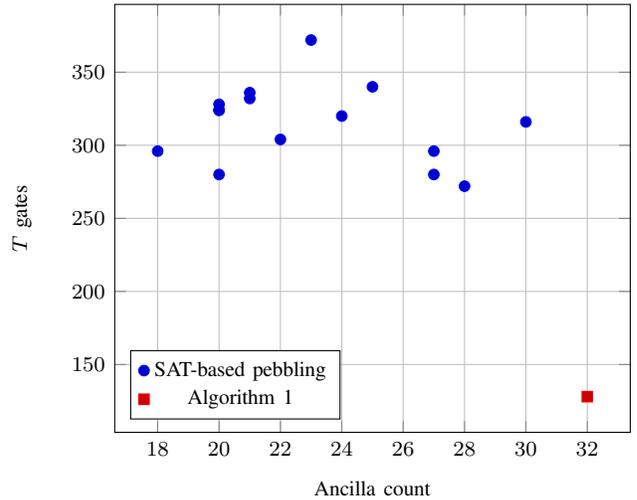
\begin{figure}[t]
  \begin{tikzpicture}[font=\footnotesize]
    \begin{axis}[legend pos=south west,xlabel={Ancilla count},ylabel={$T$ gates},grid=major]
      \plot+[only marks] coordinates {(20,324) (18,296) (21,332) (20,328) (22,304) (20,280) (21,336) (27,296) (25,340) (20,324) (23,372) (24,320) (28,272) (30,316) (27,280)};
      \plot+[only marks] coordinates {(32,128)};
      \legend{SAT-based pebbling,Algorithm~\ref{alg:heuristic}}
    \end{axis}
  \end{tikzpicture}
  \caption{Applying the SAT-based pebble strategy to the S-box benchmark.}
  \label{fig:plot}
\end{figure}
In this section, we evaluate the effect of the proposed SAT-based pebbling
strategy discussed in Section~\ref{sec:pebbling}.  We implemented the
SAT-based pebbling algorithm in Q\#~\cite{SGT+18} with Z3~\cite{MB08}
as SAT solving backend. As benchmark, we use the 8-bit S-box described
in the previous section (s8). We apply our compilation algorithm (Alg.~\ref{alg:heuristic}) to obtain a reference quantum circuit. Since its XAG representation requires 32
AND gates, using the proposed heuristic compilation algorithm we can achieve
a quantum circuit with 32 ancilla lines and 128 $T$ gates (note that 16 more qubits are required for storing inputs and outputs).  We use the
SAT-based pebbling strategy with different values for $L$ (number of
maximum ancillae) and different random seeds to obtain various other
quantum circuits with fewer number of ancilla lines.

The results are plotted in Fig.~\ref{fig:plot}.  As can be seen, the
SAT-based pebbling approach can find various different quantum circuits
with 18--30 required ancillae.  However, the $T$-count increases, as
AND gates are computed and uncomputed more than once.  Note that in
the described SAT-based pebbling strategy we do not constrain the number
of AND operations.  A weighted pebble game in which computations of
AND gates are more expensive than computations of XOR gates can help to find solutions with fewer $T$ gates, however, it comes with an
overhead in solving time.  Besides the (red) data point obtained by
the heuristic compilation algorithm, there are 3 more Pareto optimal
solutions, namely $(18,296)$, $(20,280)$, and $(28,272)$.  While all
solutions were found within a few seconds, we noticed that the
SAT-based pebbling strategy does not yet scale well to larger
benchmarks such as the EPFL benchmarks.  We plan to investigate ways to make SAT-based pebbling more scalable in future works.
\section{Future work}
In this work we propose XAGs as advantageous multi-level logic representations that allow automatic compilation to reach performances similar to manual methods. We have mostly focused on XAGs with minimal number of AND nodes, proportional to our proposed upper bound for the number of qubits and $T$-count of the final circuit. In future works, we aim at focusing on the impact of the number of XOR nodes in the graph. In fact, an XOR block of $x$ variables, requires $2 \times (x-1)$ CNOT gates to be computed and uncomputed once (see Fig.~\ref{fig:and-step}). Techniques can be borrowed from multi-level logic optimization to minimize the number of XOR nodes~\cite{FS10,BMP13} without increasing the number of ANDs and consequently minimize the number of CNOTs in the final circuit. 

In addition, it is possible to combine our method with post-synthesis CNOT optimization techniques, either heuristic~\cite{NRS+18, AAM17} or exact~\cite{Me18}, to reduce the CNOT overhead.

We present a pebbling strategy technique specifically designed to work on XAGs, where each XOR can be computed in-place, while AND nodes must be computed out-of-place. The technique can be improved by implementing a weighted pebbling game, where pebbling/unpebbling AND nodes is penalized with respect to XOR nodes. 

\section{Conclusion}

We presented a new heuristic compilation algorithm for
quantum oracles which addresses
fault-tolerant quantum computing, minimizing the $T$-count.
In our method, the number of $T$ gates depends on the number of AND gates used to represent the Boolean function $f$ as an XAG. The algorithm suggests a new upper bound on the $T$-count that is proportional to the multiplicative complexity of the function $f$, namely $4 c_{\wedge}(f)$. Future
research on multiplicative complexity will not only influence results
in cryptography, but also in quantum computing, thanks to the direct
correlation between of the multiplicative complexity of Boolean functions and
the number of $T$ gates and ancillae in the corresponding quantum circuit.

Our technique achieves better results compared to other state-of-the-art automatic compilers. In fact, these either produce too many $T$ gates, or they rely on a larger number of qubits. Nowadays, the research community is making many efforts to develop a scalable quantum technology, capable of producing quantum systems characterized by a large number of qubits. Fault tolerant quantum computing will only be possible when this technology progress is achieved. For this reason, it will be paramount to control the number of $T$ gates, rather than the number of qubits. As a consequence, compilation methods that result in large $T$-counts should be discarded in favor of low $T$-count methods.

Finally, we provide synthesis statistics for two benchmarks containing useful designs in the context of cryptography which can be useful for deriving resource cost estimates for quantum-based cryptoanalysis.

\subsubsection*{Acknowledgments}  This research was
supported by the Swiss National Science Foundation (200021-169084 MAJesty).

\bibliographystyle{IEEEtran}
\bibliography{library,extra}

\begin{thebibliography}{10}
\providecommand{\url}[1]{#1}
\csname url@samestyle\endcsname
\providecommand{\newblock}{\relax}
\providecommand{\bibinfo}[2]{#2}
\providecommand{\BIBentrySTDinterwordspacing}{\spaceskip=0pt\relax}
\providecommand{\BIBentryALTinterwordstretchfactor}{4}
\providecommand{\BIBentryALTinterwordspacing}{\spaceskip=\fontdimen2\font plus
\BIBentryALTinterwordstretchfactor\fontdimen3\font minus
  \fontdimen4\font\relax}
\providecommand{\BIBforeignlanguage}[2]{{%
\expandafter\ifx\csname l@#1\endcsname\relax
\typeout{** WARNING: IEEEtran.bst: No hyphenation pattern has been}%
\typeout{** loaded for the language `#1'. Using the pattern for}%
\typeout{** the default language instead.}%
\else
\language=\csname l@#1\endcsname
\fi
#2}}
\providecommand{\BIBdecl}{\relax}
\BIBdecl

\bibitem{Grover96}
L.~K. Grover, ``A fast quantum mechanical algorithm for database search,'' in
  \emph{Symposium on Theory and Computing}, 1996, pp. 212--219.

\bibitem{Shor97}
P.~W. Shor, ``Polynomial-time algorithms for prime factorization and discrete
  logarithms on a quantum computer,'' \emph{{SIAM} Journal on Computing},
  vol.~26, no.~5, pp. 1484--1509, 1997.

\bibitem{Harrow09}
A.~W. Harrow, A.~Hassidim, and S.~Lloyd, ``Quantum algorithm for linear systems
  of equations,'' \emph{Phys. Rev. Lett.}, vol. 103, 2009.

\bibitem{AMMR13}
M.~Amy, D.~Maslov, M.~Mosca, and M.~Roetteler, ``A meet-in-the-middle algorithm
  for fast synthesis of depth-optimal quantum circuits,'' \emph{{IEEE} Trans.
  on {CAD} of Integrated Circuits and Systems}, vol.~32, no.~6, pp. 818--830,
  2013.

\bibitem{BPP00}
J.~Boyar, R.~Peralta, and D.~Pochuev, ``On the multiplicative complexity of
  boolean functions over the basis {$(\land, \oplus, 1)$},'' \emph{Theoretical
  Computer Science}, vol. 235, no.~1, pp. 43--57, 2000.

\bibitem{Jones13}
C.~Jones, ``Low-overhead constructions for the fault-tolerant {Toffoli} gate,''
  \emph{Physical Review A}, vol.~87, no.~2, p. 022328, 2013.

\bibitem{Gidney18}
C.~Gidney, ``Halving the cost of quantum addition,'' \emph{Quantum}, vol.~2,
  p.~74, 2018.

\bibitem{Find14}
M.~G. Find, ``On the complexity of computing two nonlinearity measures,'' in
  \emph{Int'l Computer Science Symposium in Russia}, 2014, pp. 167--175.

\bibitem{BP08}
J.~Boyar and R.~Peralta, ``Tight bounds for the multiplicative complexity of
  symmetric functions,'' \emph{Theoretical Computer Science}, vol. 396, no.
  1--3, pp. 223--246, 2008.

\bibitem{testa19}
E.~Testa, M.~Soeken, L.~Amarú, and G.~De~Micheli, ``Reducing the
  multiplicative complexity in logic networks for cryptography and security
  applications,'' in \emph{DAC}, 2019.

\bibitem{BMP13}
J.~Boyar, P.~Matthews, and R.~Peralta, ``Logic minimization techniques with
  applications to cryptology,'' \emph{Journal of Cryptology}, vol.~26, no.~2,
  pp. 280--312, 2013.

\bibitem{CTP19}
{\c{C}}.~{\c{C}}alik, M.~S. Turan, and R.~Peralta, ``The multiplicative
  complexity of 6-variable {Boolean} functions,'' \emph{Cryptography and
  Communications}, vol.~11, no.~1, pp. 93--107, 2019.

\bibitem{Bennett89}
C.~H. Bennett, ``Time/space trade-offs for reversible computation,''
  \emph{{SIAM} Journal on Computing}, vol.~18, no.~4, pp. 766--776, 1989.

\bibitem{SRWM17b}
M.~Soeken, M.~Roetteler, N.~Wiebe, and G.~De~Micheli, ``Hierarchical reversible
  logic synthesis using {LUTs},'' in \emph{Design Automation Conference}, 2017,
  pp. 78:1--78:6.

\bibitem{Meuli18}
G.~Meuli, M.~Soeken, M.~Roetteler, N.~Wiebe, and G.~De~Micheli, ``A best-fit
  mapping algorithm to facilitate {ESOP}-decomposition in {Clifford+ T} quantum
  network synthesis,'' in \emph{Asia and South Pacific Design Automation
  Conference}.\hskip 1em plus 0.5em minus 0.4em\relax IEEE Press, 2018, pp.
  664--669.

\bibitem{BK05}
S.~Bravyi and A.~Kitaev, ``Universal quantum computation with ideal {Clifford}
  gates and noisy ancillas,'' \emph{Physical Review A}, vol.~71, p. 022316,
  2005.

\bibitem{OGC17}
J.~O'Gorman and E.~T. Campbell, ``Quantum computation with realistic
  magic-state factories,'' \emph{Physical Review A}, vol.~95, no.~3, p. 032338,
  2017.

\bibitem{Meuli19}
G.~Meuli, M.~Soeken, M.~Roetteler, N.~Bjorner, and G.~De~Micheli, ``Reversible
  pebbling game for quantum memory management,'' in \emph{Design, Automation
  and Test in Europe}, 2019.

\bibitem{GC99}
D.~Gottesman and I.~L. Chuang, ``Quantum teleportation is a universal
  computational primitive,'' \emph{Nature}, vol. 402, pp. 390--393, 1999.

\bibitem{HC17}
M.~Howard and E.~Campbell, ``Application of a resource theory for magic states
  to fault-tolerant quantum computing,'' \emph{Phys. Rev. Lett.}, vol. 118,
  2017.

\bibitem{PRS17}
A.~Parent, M.~Roetteler, and K.~M. Svore, ``{REVS:} {A} tool for
  space-optimized reversible circuit synthesis,'' in \emph{Int'l Conf. on
  Reversible Computation}, 2017, pp. 90--101.

\bibitem{Knuth4f6}
D.~E. Knuth, \emph{The Art of Computer Programming, Volume 4, Fascicle 6:
  Satisfiability}.\hskip 1em plus 0.5em minus 0.4em\relax Addison-Wesley, 2015.

\bibitem{Mishchenko06}
A.~{Mishchenko}, S.~{Chatterjee}, and R.~{Brayton}, ``{DAG}-aware {AIG}
  rewriting: a fresh look at combinational logic synthesis,'' in \emph{Design
  Automation Conference}, 2006.

\bibitem{fazel07}
K.~Fazel, M.~A. Thornton, and J.~Rice, ``{ESOP}-based toffoli gate cascade
  generation,'' in \emph{Pacific Rim Conference on Communications, Computers
  and Signal Processing}, 2007.

\bibitem{grassl16}
M.~Grassl, B.~Langenberg, M.~Roetteler, and R.~Steinwandt, ``Applying
  grover’s algorithm to {AES}: quantum resource estimates,'' in
  \emph{Post-Quantum Cryptography}, 2016.

\bibitem{SGT+18}
K.~Svore, A.~Geller, M.~Troyer, J.~Azariah, C.~Granade, B.~Heim,
  V.~Kliuchnikov, M.~Mykhailova, A.~Paz, and M.~Roetteler, ``{Q\#}: Enabling
  scalable quantum computing and development with a high-level {DSL},'' in
  \emph{Real World Domain Specific Languages Workshop}, 2018, pp. 7:1--7:10.

\bibitem{MB08}
L.~M. de~Moura and N.~Bj{\o}rner, ``{Z3:} an efficient {SMT} solver,'' in
  \emph{Int'l Conf. on Tools and Algorithms for the Construction and Analysis
  of Systems}, 2008, pp. 337--340.

\bibitem{FS10}
C.~Fuhs and P.~Schneider{-}Kamp, ``Synthesizing shortest linear straight-line
  programs over {GF(2)} using {SAT},'' in \emph{Int'l Conf. on Theory and
  Applications of Satisfiability Testing}, 2010, pp. 71--84.

\bibitem{NRS+18}
Y.~Nam, N.~J. Ross, Y.~Su, A.~M. Childs, and D.~Maslov, ``Automated
  optimization of large quantum circuits with continuous parameters,''
  \emph{npj Quantum Information}, vol.~4, no.~23, pp. 1--12, 2018.

\bibitem{AAM17}
M.~Amy, P.~Azimzadeh, and M.~Mosca, ``On the {CNOT}-complexity of {CNOT}-phase
  circuits,'' \emph{arXiv preprint arXiv:1712.01859}, 2017.

\bibitem{Me18}
G.~Meuli, M.~Soeken, and G.~De~Micheli, ``{SAT}-based $\{$CNOT, T$\}$ quantum
  circuit synthesis,'' in \emph{Int'l Conf. on Reversible Computation}.\hskip
  1em plus 0.5em minus 0.4em\relax Springer, 2018, pp. 175--188.

\end{thebibliography}

\end{document}